**Low-Temperature Synthesis of Stable $CaZn_2P_2$ Zintl Phosphide Thin Films as Candidate Top Absorbers**

Shaham Quadir[1*], Zhenkun Yuan[2], Guillermo Esparza[3], Sita Dugu[1], John Mangum[1], Andrew Pike[2], Muhammad Rubaiat Hasan[4], Gideon Kassa[2], Xiaoxin Wang[2], Yagmur Coban[2], Jifeng Liu[2], Kirill Kovnir[4,5], David P. Fenning[3], Obadiah G. Reid[1,6], Andriy Zakutayev[1], Geoffroy Hautier[2], Sage R. Bauers[1†]

[1]Materials Science Center, National Renewable Energy Laboratory, Golden, CO 80401, USA

[2]Thayer School of Engineering, Dartmouth College, Hanover, NH 03755, USA

[3] Department of Chemical and Nano Engineering, University of California, San Diego, La Jolla, CA 92093, USA

[4] Department of Chemistry, Iowa State University, Ames, IA 50011, USA

[5] Ames National Laboratory, U.S. Department of Energy, Ames, IA 50011, USA

[6] Renewable and Sustainable Energy Institute, University of Colorado, Boulder, Boulder, CO 80309, USA

---

[*]shaham.quadir@nrel.gov

[†]sage.bauers@nrel.gov

**Abstract**

The development of tandem photovoltaics and photoelectrochemical solar cells requires new absorber materials with band gaps in the range of ~1.5–2.3 eV, for use in the top cell paired with a narrower-gap bottom cell. An outstanding challenge is finding materials with suitable optoelectronic and defect properties, good operational stability, and synthesis conditions that preserve underlying device layers. This study demonstrates the Zintl phosphide compound $CaZn_2P_2$ as a compelling candidate semiconductor for these applications. We prepare phase pure, ~500 nm-thick $CaZn_2P_2$ thin films using a scalable reactive sputter deposition process at growth temperatures as low as 100 °C, which is desirable for device integration. UV-vis spectroscopy shows that $CaZn_2P_2$ films exhibit an optical absorptivity of ~$10^4$ $cm^{-1}$ at ~1.95 eV direct band gap. Room-temperature photoluminescence (PL) measurements show near-band-edge optical emission, and time-resolved microwave conductivity (TRMC) measurements indicate a photoexcited carrier lifetime of ~30 ns. $CaZn_2P_2$ is highly stable in both ambient conditions and moisture, as evidenced by PL and TRMC measurements. Experimental data are supported by first-principles calculations, which indicate the absence of low-formation-energy, deep intrinsic defects. Overall, our study should motivate future work integrating this potential top cell absorber material into tandem solar cells.

## Introduction

The growing demand for sustainable and clean energy sources has propelled extensive research in photovoltaic (PV) materials and devices aimed at increasing solar energy conversion efficiency.[1-4] In recent decades, most of these efforts have focused on single-junction thin-film architectures, due to their potential for becoming low-cost and high-performance technologies.[2, 5] Among thin-film solar absorbers, CdTe, GaAs, and $Cu(In,Ga)Se_2$ have been widely studied, with single-junction device efficiencies reaching nearly 24% [6-9] Thin-film solar cells based on emerging inorganic absorber materials, such as SnS, $Cu_2ZnSn(S,Se)_4$, and $Sb_2(S,Se)_3$, have also made significant progress, but currently their efficiency is limited due to various bulk and interface issues.[10-15] Meanwhile, perovskite solar cells, based on, e.g., $CH_3NH_3PbI_3$, have made remarkable progress, with the efficiencies rapidly exceeding 25%, attributed to their tunable band gaps, large absorption coefficients, and high defect tolerance.[16, 17] Despite the advances in thin-film technologies, single-junction crystalline Si solar cells dominate today's global PV market. However, in the form of single-junction devices, both crystalline Si and thin-film solar cells will be restricted by the detailed-balance limit, which predicts a theoretical maximum power conversion efficiency (PCE) of ~33% for single-junction solar cells.[18-20]

An effective method to go beyond the single-junction detailed-balance limit is to introduce tandem structures comprising a top cell with a high band gap (1.5–2.3 eV) absorber and a bottom cell based on a well matched lower band gap absorber, which better utilizes the solar spectrum.[21-23] Theoretically, the PCE limit of such stacked architectures could increase up to ~47%, well surpassing the detailed-balance limit for single-junction devices.[24-26] Various absorber pairs, most of which are based on established bottom-cell materials, are under intensive development.[27] Lead halide perovskite top cells combined with Si bottom cells are the most-studied tandem devices, with the highest reported efficiency being 33.9%.[28, 29] However, there are ongoing efforts to explore alternative materials beyond lead halide perovskites due to ongoing challenges with long-term stability.[30, 31] III–V (GaInP) and II-VI (CdTe) semiconductors exhibit promising performance as top cell absorbers,[23, 32] but the high manufacturing costs of III–V solar cells and large thermal budget currently required for high-quality CdTe pose significant challenges for their implementation in tandem devices.[33, 34] Emerging materials, such as chalcogenide perovskites (e.g., $BaZrS_3$), are also being considered as top-cell absorbers because they are highly stable and possess excellent baseline properties. However, high-temperature synthesis and problematic defect chemistries have complicated deployment.[35-37] Therefore, it is necessary to identify new solar absorbers that combine superior optoelectronic and defect properties with operational stability, abundance in earth's crust, and synthesizability under mild conditions.

Recently, materials screening based on high-throughput first-principles calculations has enabled researchers to rapidly discover new candidate solar absorbers.[38-46] Defect-related properties, and especially defect-induced nonradiative carrier recombination, have started to be included in the computational screening with the goal to identify "defect-tolerant" materials.[44-46] Using this state-of-the-art computational screening approach, we recently identified the Zintl phosphide $BaCd_2P_2$ in the $P\bar{3}m1$ structure ($CaAl_2Si_2$ prototype) as an attractive thin-film solar absorber.[46] Unoptimized $BaCd_2P_2$ powder was experimentally shown to have a long carrier lifetime of up to 30 ns, which was only surpassed by CdTe after decades of optimization.[6] $BaCd_2P_2$ was also found to be stable, in terms of both structural and optoelectronic properties, under various harsh thermal and chemical treatments. While clearly promising, measurements on $BaCd_2P_2$ were made on powder samples unsuitable for devices, and the 1.45 eV band gap, while suitable for single-junction photovoltaics, is too low for tandem top cell applications.

$BaCd_2P_2$ is just one of the materials in a large family of $AM_2P_2$ Zintl compounds where $A$ = Ca, Sr, Ba and $M$ = Zn or Cd.[47-49] Inspired by $BaCd_2P_2$, these related $AM_2P_2$ materials have attracted our interest as potential tandem top cell absorbers. Using the $AM_2P_2$ materials as top cell absorbers and based on their calculated band gaps (see **Table S1**) we find that an ideal PCE of >40% could be achieved by pairing the $AM_2P_2$ top cells with a suitable lower band gap bottom cell such as Si or CIGS (**Figure 1**). This calculation is based on the detailed-balance limit and does not consider current matching or reflective losses and thus represents an ideal four-terminal tandem configuration.[50] For materials with $M$ = Zn, namely $CaZn_2P_2$ and $SrZn_2P_2$, the fundamental band gap is indirect. In this case both the fundamental (dotted lines) and direct band gaps (dashed lines) are shown. Among these $AM_2P_2$ materials, $CaZn_2P_2$ is especially appealing with a computed band gap of 1.55 eV (fundamental) and 1.89 eV (direct). We note that $CaZn_2P_2$ has been suggested as a photoelectrode for water splitting devices,[51] and that Katsube et al. synthesized bulk $CaZn_2P_2$ crystals measuring a direct band gap of $E_g$ = 2.05 eV, in reasonable agreement with the calculated band gap.[52]

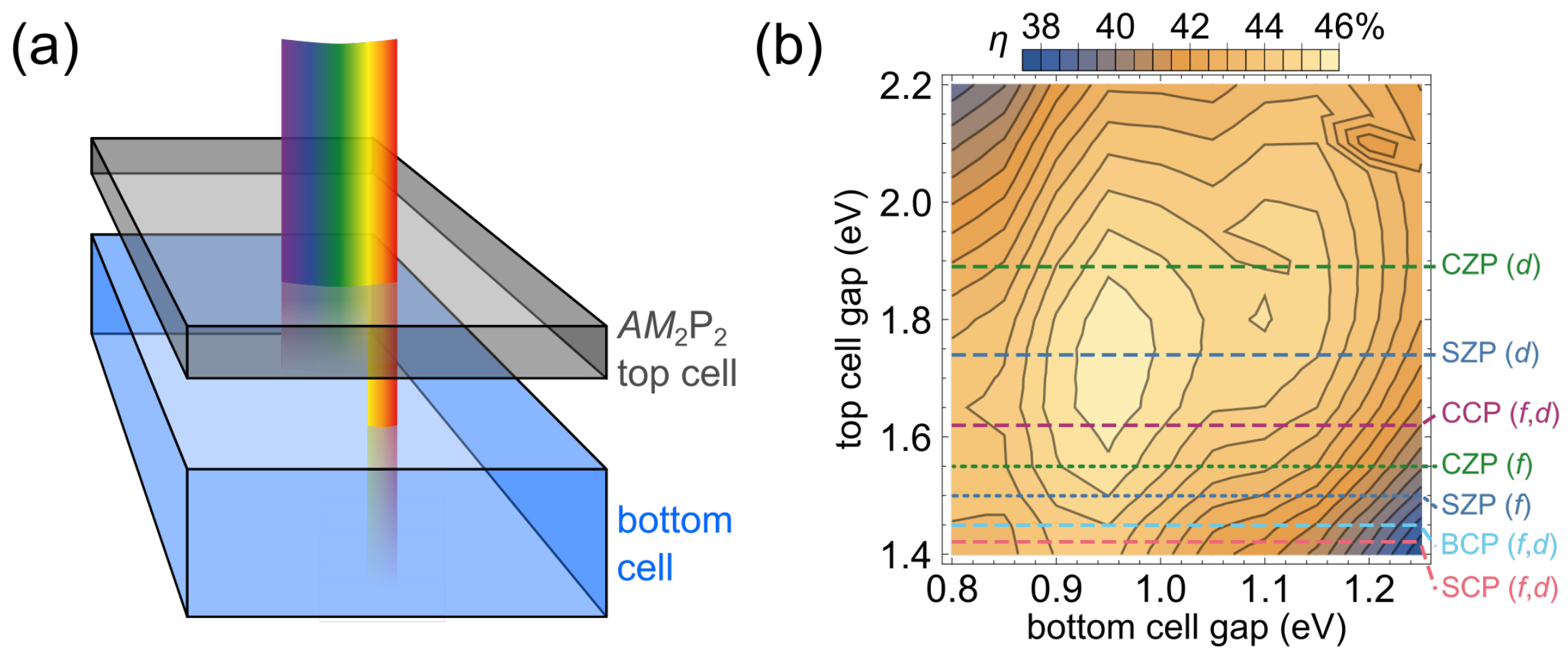

**Figure 1.** $AM_2P_2$ compounds as candidate tandem solar cell materials. (a) Schematic of $AM_2P_2$ as a tandem top cell. (b) Theoretical device efficiency for tandem structures using $AM_2P_2$ top cells, calculated based on the detailed balance limit. For the $AM_2P_2$ compound labels, CZP = $CaZn_2P_2$, SZP = $SrZn_2P_2$, CCP = $CaCd_2P_2$, BCP = $BaCd_2P_2$, and SCP = $SrCd_2P_2$; $f$ denotes a fundamental band gap and $d$ denotes a direct band gap.

In this article, we report a synthesis route for $CaZn_2P_2$ Zintl-phosphide thin films as candidate top cell absorbers for tandem solar cells, achieving crystalline films at growth temperatures ($T_{growth}$) as low as ~100 °C.

Using a scalable reactive sputtering technique from simple metallic precursors and $PH_3$ gas, we prepare phase-pure $CaZn_2P_2$ in the known $P\bar{3}m1$ structure. The uniform and compact films exhibit both photoluminescence (PL) near the 1.95 eV direct band gap and high optical absorption of $10^4$ at ~1.95 eV and $10^5$ $cm^{-1}$ above 2.6 eV, thus meeting a few preliminary criteria for a top cell PV absorber. Time-resolved microwave conductivity (TRMC) measurements reveal a carrier lifetime of up to 30 ns at low laser fluence. Finally, first-principles calculations rationalize the measured long carrier lifetime, showing the absence of low-formation-energy, deep intrinsic defects. These combined experimental and theoretical results for $CaZn_2P_2$ provide insight into the material's fundamental properties as well as its practical applicability as a top cell absorber for tandem solar cells.

## Results

*Growth and structural characterization*

**Figure 2a** shows X-ray diffraction (XRD) patterns as heatmaps for three combinatorial films grown with 100 °C ≤ $T_{growth}$ ≤ 300 °C. As seen by comparing with the grey simulated traces, for each $T_{growth}$ the diffraction peaks are well matched with the $P\bar{3}m1$ trigonal lattice that was previously reported for $CaZn_2P_2$ (**Figure 2b**) over a broad Ca/Zn composition range, with no crystalline secondary phases.[47] In this structure, Ca is octahedrally coordinated by P and Zn is tetrahedrally coordinated by P and the cations arrange themselves into layers. Overall, $CaZn_2P_2$ can be described as having $Ca^{2+}$ cations with a $(Zn_2P_2)^{2-}$ polyanion. While such Zintl compounds are well known in some energy materials, such as thermoelectrics,[53] they are not often used in solar absorption. The relative peak intensities change both with $T_{growth}$ and composition, which is likely a convolution of crystallographic texture and antisite defect (or vacancy) formation. The overall peak intensity increases with $T_{growth}$, suggesting an increase in crystallinity, but crystalline, phase-pure $CaZn_2P_2$ forms even at very low $T_{growth}$ of 100 °C. On the other hand, bulk $CaZn_2P_2$ requires significantly higher temperatures to synthesize; whether this was due to large bond formation barriers or the need to drive solid state diffusion was unknown. Our bulk synthesis exploration showed that at temperatures lower than 1000°C admixture of $Zn_3P_2$ and presumably binary Ca-P are present in the samples pointing towards solid-state diffusion limitations. Conversely, when the process is no longer limited by slow diffusion, such as in thin film growth from vapor, $CaZn_2P_2$ readily forms at low temperature.

To better understand composition variations in off-stoichiometric $CaZn_2P_2$, we increased the relative Ca flux to the substrate while maintaining all other deposition conditions. We found that Ca-rich conditions (overall Ca fraction of ~0.3–0.4, where 0.2 is stoichiometric $CaZn_2P_2$) make the film amorphous (**Figure S1**). Ca and Zn are both nominally $2^+$ cations and sputtering often affords significant amounts of cation antisite disorder in ternary pnictides[54], so the amorphization likely arises from the extreme size mismatch when trying to force Ca, which prefers a high coordination number (Shannon radius 1Å when octahedral), into a tetrahedral

Zn site (Shannon radius 0.6Å).[55] This observation generally supports that there will be high cation antisite defect energies in $CaZn_2P_2$ and related materials.

At $T_{growth}$ = 400 °C, only visually transparent films were produced. Presumably only a thin layer of Ca was deposited, because of the high vapor pressure of Zn and P, and this subsequently oxidized into CaO during an $O_2$-purge procedure used to clear the growth chamber of residual $PH_3$ prior to moving samples into the load lock. Some desorption of Zn, the most volatile element in this system,[56] already occurs at $T_{growth}$ = 300 °C, as evidenced in **Figure 2c** by the film's composition moving toward a more Ca- and P-rich regime. Because P is also volatile, the slight increase in P concentration may be indicative of increased thermal cracking efficiency of $PH_3$ gas into elemental P at the substrate surface with the increased $T_{growth}$.[57]

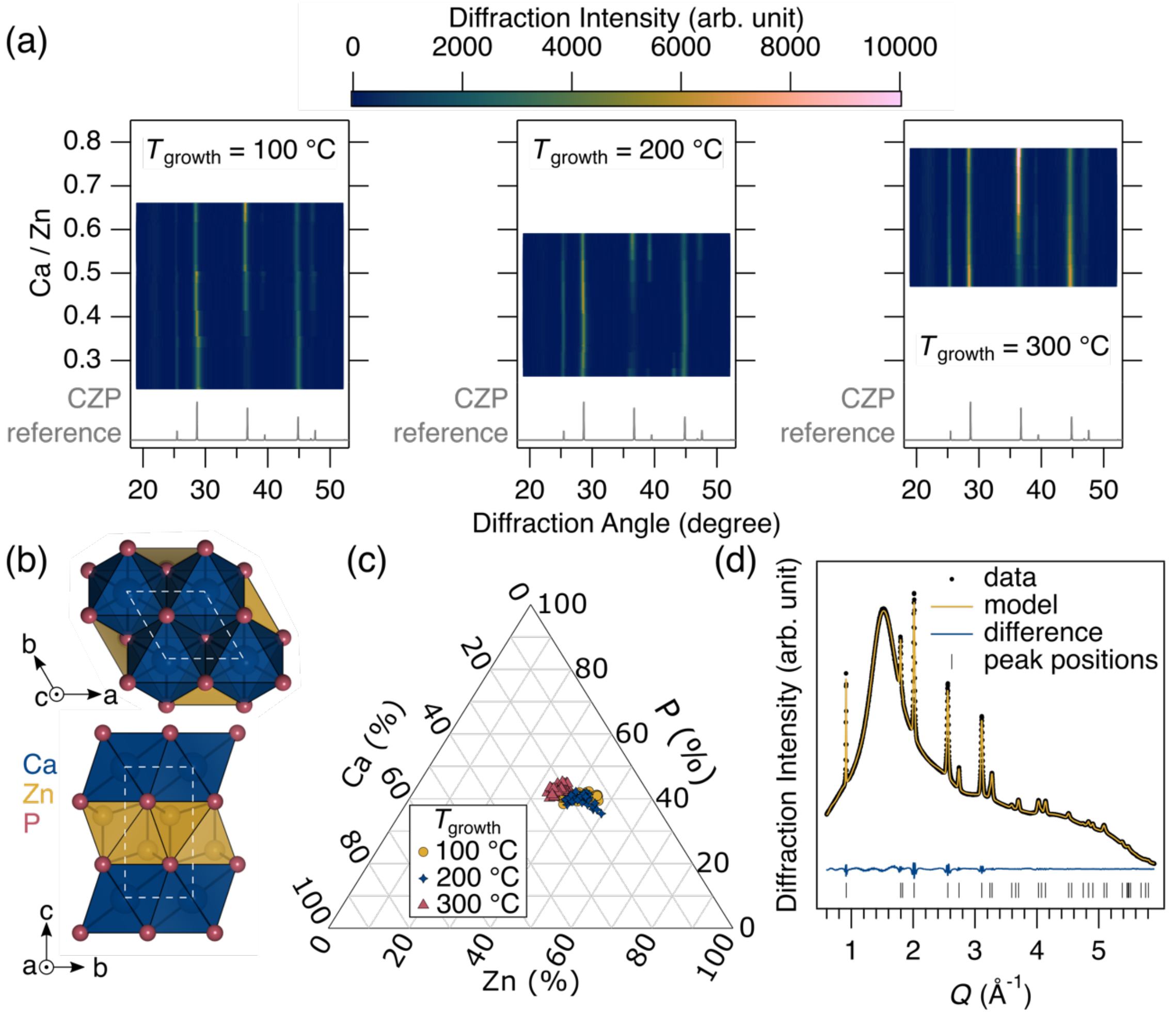

**Figure 2.** Structure and composition of $CaZn_2P_2$ thin films. (a) X-ray diffraction heatmaps for $CaZn_2P_2$ films grown with different compositions and growth temperatures ($T_{growth}$). Films grown at all conditions exhibit diffraction peaks exclusively from the reference phase.[47] (b) Crystal structure schematics for $CaZn_2P_2$. (c) Ternary phase diagram showing the full compositions of films presented in (a). (d)

Integrated wide angle x-ray scattering pattern for a uniform, stoichiometric $CaZn_2P_2$ film along with a LeBail fit to the data.

All further films were prepared at 200 °C because, in concert with our other deposition conditions, this $T_{growth}$ resulted in the best combination of crystallinity, matching the reference XRD patterns, and stoichiometric compositions. However, we note that the combinatorial composition gradients found in these films are small relative to other material systems grown using similar growth approaches and similar volatile chemistries.[58] This suggests an adsorption-controlled growth mode—where composition is pinned by the Ca flux so long as an over flux of Zn and P is provided—could be achieved at higher $T_{growth}$. This could lead to very high-quality $CaZn_2P_2$ , in similar fashion to, e.g., CIGS, where excess Se flux is provided.[59]

A uniform, stoichiometric, and phase pure $CaZn_2P_2$ thin film was prepared by rotating the substrate. This film was characterized by synchrotron grazing incidence wide angle X-ray scattering (GI-WAXS) and integrated to generate a powder diffraction pattern. The experimental data and a LeBail whole pattern refinement against the reported $P\overline{3}m1$ structure is presented in **Figure 2d.** The fit returns lattice parameters of $a$ = 4.0333 Å and $c$ = 6.898 Å, and a low overall weighted profile R-factor of 1.43%, in excellent agreement with prior reports on bulk $CaZn_2P_2$.[47] There is a small degree of crystallographic texture, as evidenced by nonuniform Debye ring intensity in the raw detector image (**Figure S2**), which precludes a full structural refinement using the Rietveld method. The broad signal in the diffraction pattern centered at Q ≈ 1.5 ($Å^{-1}$) is from the a-$SiO_2$ substrate.

To compare against our thin films, which we believe are the first reported for $CaZn_2P_2$, we also prepared high quality $CaZn_2P_2$ powder by solid state reaction from elements. While the diffraction patterns are qualitatively similar between film and bulk (**Figure S3**), there are a couple of features that stand out. First, a small peak shift to lower $Q$ is observed in the films. Refining the bulk powder pattern yields lattice parameters of $a$ = 4.038 Å and $c$ = 6.836 Å. The origin of the slight discrepancy remains to be determined. Second, the full width at half maximum of diffraction peaks from the thin film are about double the powder's, suggesting small crystallites.

The crystalline coherence length, $L$, of the $CaZn_2P_2$ film was estimated using the Scherrer Equation ($L = \frac{k\lambda}{\beta\cos(\theta)}$), which returned a relatively small size of ~25 nm. This is of similar order to apparent grains observed in plan-view scanning electron microscopy (SEM) for films grown on a-$SiO_2$, but smaller than the apparent grains for films grown on FTO-coated glass substrates, observed in cross-sectional SEM, as depicted in **Figure S4**. This image shows ~50–100 nm grains creating a compacted film with a thickness close to 450 nm. Such grain sizes are small compared to those found in CZTS, CdTe, and perovskite absorbers.[60, 61] However, small crystallites are typical for sputtered thin films, especially when grown at low temperature. Increasing grain size and minimizing the impact of extended defects at the grain boundaries will be an exciting challenge for the community as $CaZn_2P_2$ and related materials mature.

Further insight into the structural properties of sputtered $CaZn_2P_2$ thin films was gained through high-resolution transmission electron microscopy (HR-TEM) analysis, carried out on a specimen prepared by focused ion beam (FIB) milling. **Figure 3** presents results from the HR-TEM characterization. **Figure 3a** shows a bright-field TEM micrograph of the entire thickness of the film showing diffraction contrast that highlights the columnar microstructure of the film. Most of the film comprises vertical grains that are ca. 30 nm wide; there appears to be a ~30–50 nm thick amorphous layer near the substrate. Selected area electron diffraction rings and extracted *d*-spacings, shown in **Figure 3b**, can be indexed to the $CaZn_2P_2$ phase.

Scanning TEM energy-dispersive X-ray spectroscopy (STEM-EDS) collected throughout the film's thickness showed Zn, Ca, and P, as expected as well as O, which was more prevalent at the front surface and back interface (**Figure 3c**). The O at the back interface is expected from the substrate's native oxide; O at the front surface is also not surprising, given the oxyphilic cations in $CaZn_2P_2$. Quantified EDS compositions as a function of depth are shown in **Figure 3d**. Considering only the $CaZn_2P_2$ constituents, the overall composition matches well to the $CaZn_2P_2$ stoichiometry. However, a Ca-rich region is observed near the substrate, providing an explanation for the amorphous layer observed in **Figure 3a**. About 10% C and O are found by EDS, and their intensities appear correlated with Ca, supporting the formation of Ca-carbonates. In **Figure S5** and **Figure S6** we employ X-ray photoelectron spectroscopy (XPS) to show that C is adventitious and does not penetrate deeply into the film, confirming that the C and O measured through the thickness most-likely arises from surface adsorbates after the lamella was prepared.

A high-resolution bright-field STEM image collected from a crystalline region of the film is shown in **Figure 3e**. On the left side of the image, a grain boundary is observed; the change of observable crystalline facets is abrupt, suggesting that there is little amorphous material at the grain boundaries. The striated region on the right hand of the image can be integrated to reveal that the stripes belong to Ca and Zn planes perpendicular to the [00*l*] lattice vector of $CaZn_2P_2$, as shown in **Figure 3f**, where the *c*-lattice parameter measured by GI-WAXS matches well against the brightness variations measured in HR-STEM.

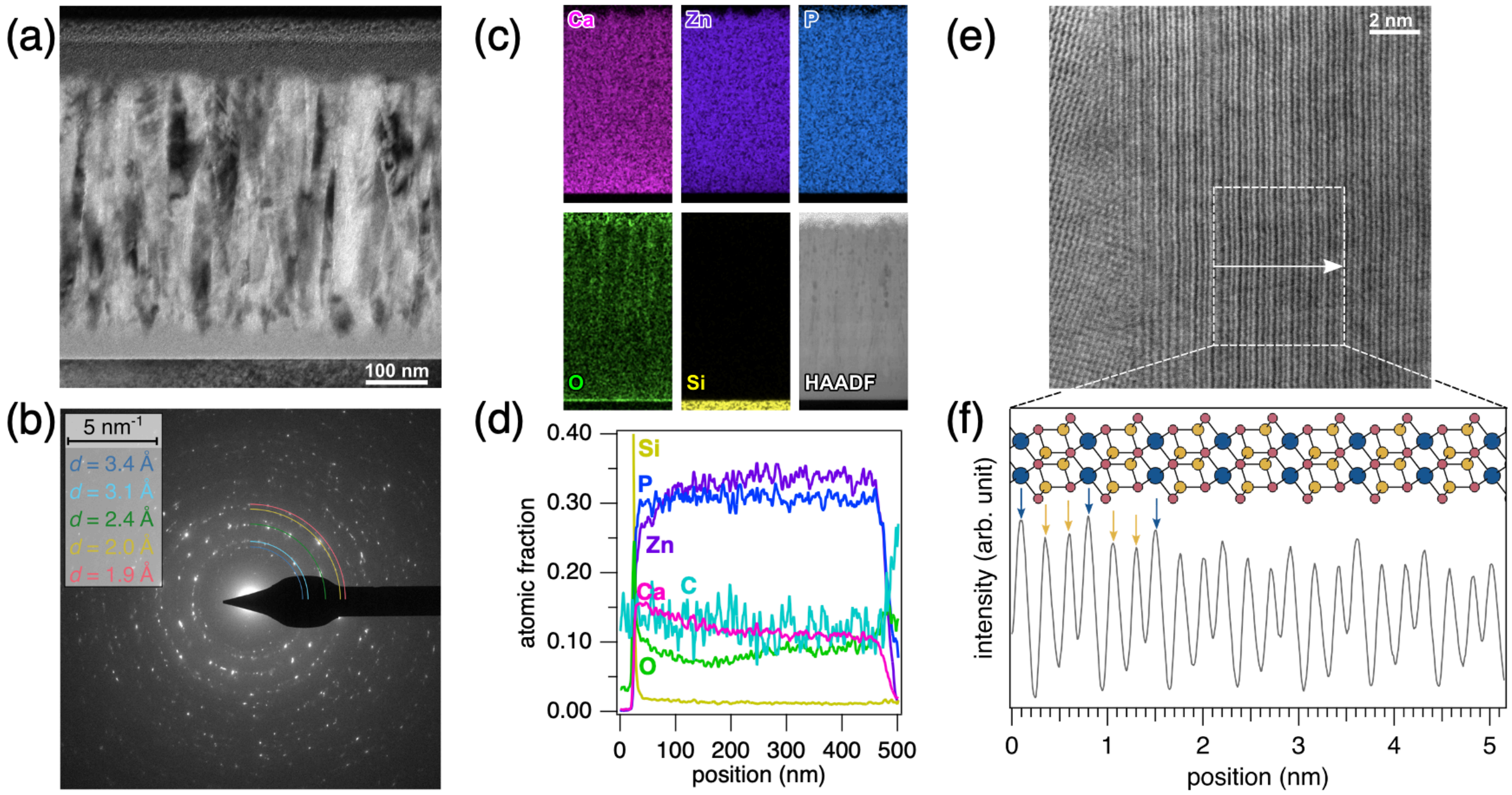

**Figure 3.** Electron microscopy from a $CaZn_2P_2$ film grown on Si with native oxide. (a) Bright-field TEM micrograph of the film showing diffraction contrast from columnar crystallites. (b) Selected area electron diffraction indexed to the $CaZn_2P_2$ phase. (c) and (d) Energy-dispersive X-ray spectroscopy (c) elemental maps and (d) quantified line scans. (e) High-resolution bright-field STEM micrograph and (f) integrated line intensity compared against the $CaZn_2P_2$ crystal structure.

*Optical Characterization*

The optical reflection ($R$) and transmission ($T$) of white light, measured by ultraviolet- and visible-range spectroscopy (UV-vis), through a $CaZn_2P_2$ film are shown as a function of photon energy ($E_{ph}$) in **Figure 4a**. The $T$ and $R$ data exhibit an oscillatory below and near the band edge due to the multiple reflections from the front and back interfaces of the ~500 nm thick film, which further proves the uniformity and smoothness of the $CaZn_2P_2$ films. The transmittance and reflectance spectra are fit numerically using transfer matrix method [62] in order to take into account the interference effect in the $CaZn_2P_2$ films, which yields wavelength-dependent refractive index $n$ and extinction coefficient $k$ for comparison with theoretical predictions. Details of the fitting method can be found the Supporting Information of Ref. [63]. As shown in **Figure 4a**, the fitting agrees very well with the experimental $T$ and $R$ data, except for a higher reflectance at $E_{\mathrm{ph}} \gtrsim 2.2$ eV than the experiment. Considering that the positions of reflectance peaks and valleys are still in very good agreement at $E_{\mathrm{ph}} \gtrsim 2.2$ eV, the refractive index from the fitting still matches the thin film interference pattern. Therefore, a possible reason for lower reflectance values measured experimentally is the light trapping effect due to multiple scattering at shorter wavelengths induced by the nanostructured grains in the $CaZn_2P_2$ thin films shown in **Figure 3a**, since photons at shorter wavelengths are more sensitive to scattering. This is a beneficial feature for solar absorbers if the nanostructures do not otherwise harm other properties (e.g., transport). The $n$ and $k$ from the fitting are compared with the theoretical modeling in **Figure S7** in the **Supporting Information**. Overall, the theoretical model shows a slight blueshift at high photon energies, but the key features and values of $n$ and $k$ are in good agreement with the $n$, $k$ from the fitting of experimental data. The absorption coefficient ($\alpha$), shown in **Figure 4b**, was determined from the extinction coefficient $k$ using the relation $\alpha = 4\pi k/\lambda$, and compared to the theoretical modeling. An indirect gap transition at ~1.6 eV and a direct gap absorption edge at ~1.95 eV is observed in the absorption spectrum, in good agreement with the theoretically modelled bandgaps (1.55 eV indirect gap and 1.89 eV direct gap). Note that the theoretical model in **Figure 4b** does not consider the indirect gap absorption, which can be significant at photon energies well above the indirect gap, as is the case for Si. Therefore, the absolute values of the theoretical prediction are lower than the experiment. The magnitude of $\alpha$ shows high values exceeding $10^5$ $cm^{-1}$ in the visible spectral region for efficient solar absorption. It should be emphasized that all the other practical thin-film solar cell absorbers, including InP, CdTe, CZTS, and perovskite, show a similar band-edge α of $10^4 - 10^5$ $cm^{-1}$.[64, 65] Overall, the shape and magnitude of the experimentally measured $\alpha$ curve matches reasonably well against the theoretical trace, especially given the excellent match of the fundamental/indirect and direct gaps. Inset into **Figure 4b** is a photograph of a combinatorial $CaZn_2P_2$ film, where an optical change can be observed along the composition gradient, suggesting a measure of band gap tunability might be possible. $\alpha$ is directly measured from UV-vis data for 11 points along this film and shown in **Figure S8**.

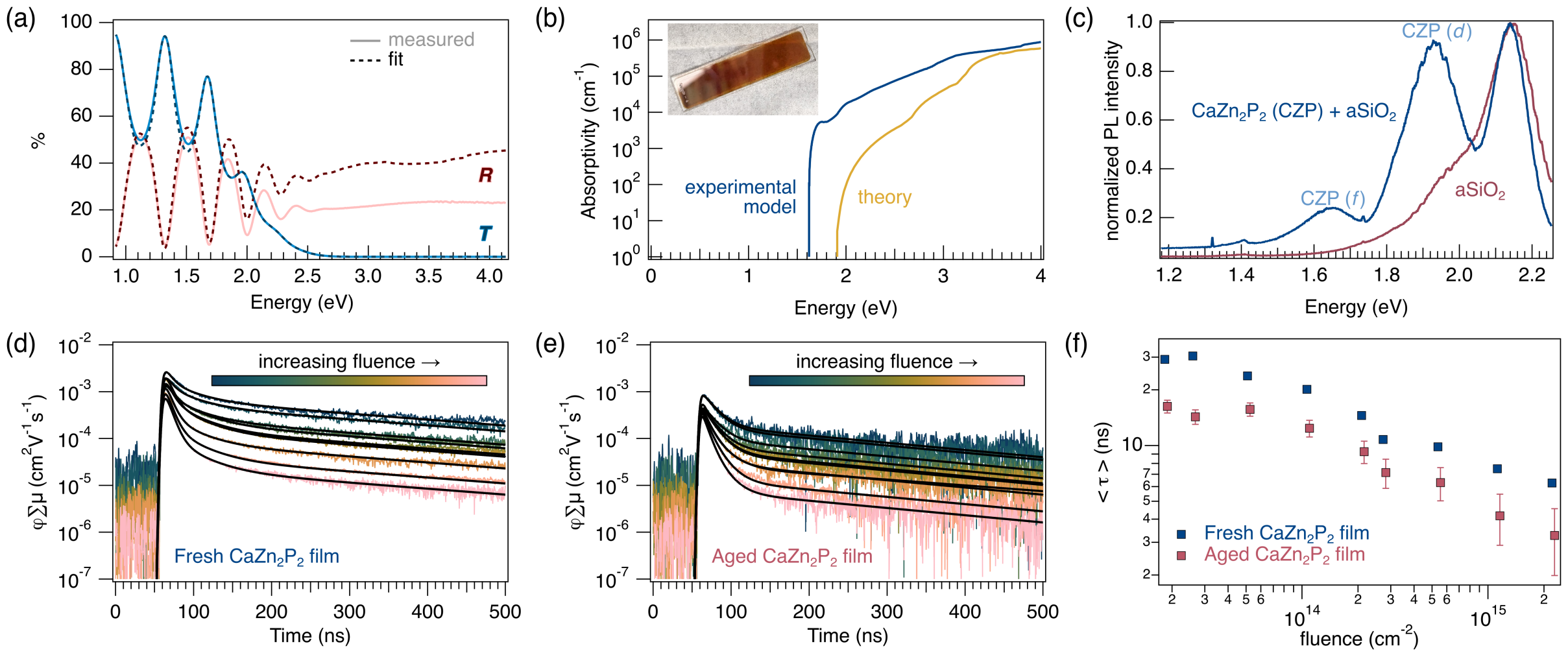

**Figure 4.** Optical properties of $CaZn_2P_2$ films. (a) Reflection ($R$), and transmission ($T$) collected by UV-vis spectroscopy. (b) Experimentally modelled and theoretically calculated absorption coefficient ($\alpha$) for $CaZn_2P_2$. Inset is a photograph of a compositionally combinatorial $CaZn_2P_2$ film. (c) Photoluminescence of $CaZn_2P_2$ reveals indirect ($f$) and direct ($d$) transition ~1.60 and 1.95 eV respectively (red curve represents the bare aSiO2 substrate) (d) and (e) Time resolved microwave conductivity (TRMC) transients collected from (d) a film that was < 2 days old and (e) a film aged in atmosphere for >2 months prior to measuring. (f) Carrier lifetimes as a function of fluence extracted from TRMC data.

The optical properties of the $CaZn_2P_2$ thin-film absorbers were studied by room-temperature photoluminescence (PL) measurements as shown in **Figure 4c**. The data from the a$SiO_2$ substrate is also presented for comparison. The peak at 2.15 eV is clearly due to the color center in the silica substrate, while the stronger peak at 1.95 eV and the weaker peak at 1.65 eV correspond very well to the calculated direct bandgap of 1.89 eV and the fundamental gap of 1.55 eV, respectively, and also agree with the absorption edges observed in **Figure 4b**. To test the stability of $CaZn_2P_2$'s optical properties under moisture, we immersed a film in water for ~90 seconds and repeated the PL measurement. We found that $CaZn_2P_2$ retained >90% of its initial PL intensity **(Figure S9)** after water soaking, which is within the measurement-to-measurement variability due to changes in location on the sample, focusing, laser condition, etc.

To explore the dynamics of optically excited charge carriers with respect to the $CaZn_2P_2$ film's stability, we performed laser fluence dependent time resolved microwave conductivity (TRMC) measurements. One freshly prepared $CaZn_2P_2$ film and another 2-month-old film aged in ambient indoor air were used for this experiment. The films were excited with a 530 nm wavelength laser at different intensities ($10^{12}$ – $10^{15}$ photons pulse$^{-1}$ cm$^{-2}$). The transients were fitted using a sum of three exponential functions numerically convolved with the instrumental response, and the sum of the pre-exponential factors was used to estimate the maximum yield-mobility product.[66] Assuming each of the absorbed photons creates a free carrier pair that contributes to microwave conductivity, we can estimate the sum of the mobilities of the individual carriers using the equation given below [67, 68],

$$\varphi \sum \mu \propto \frac{\Delta G}{I_0 F_{\mathrm{A}}}$$

where $\varphi$ is the charge carrier generation yield (e.g., unity), $\Sigma\mu$ is the sum of electron and hole mobilities, $\Delta G$ is the change in microwave conductance, $I_0$ is the incident intensity per pulse, and $F_{\mathrm{A}}$ is the fraction of incident photons absorbed within the sample. Buried in the proportionality factor are a number of universal and geometric parameters; this factor can be readily determined through electromagnetic simulations.[69]

For both films, TRMC transients were measured between ~$10^{12}$ – $10^{13}$ photons pulse$^{-1}$ cm$^{-2}$, as shown in **Figures 4d and 4e**. The effective mobility at the lowest measured fluence is around $(7\pm2)\times10^{-3}$ cm$^2$ V$^{-1}$ s$^{-1}$ for the freshly prepared sample and $(3 \pm 1)\times10^{-3}$ cm$^2$ V$^{-1}$ s$^{-1}$ for the aged sample. However, these numbers are very likely to be limited by the small crystalline size of the present samples (~25 nm for films grown on a-$SiO_2$).[70] For the fresh sample at low fluence, carrier lifetime on the order of 30 ns is observed (**Figure 4f**), and this only slightly decreases for the aged sample. As the laser excitation power increases, carrier lifetimes become shorter, and there is a concomitant reduction of TRMC signal magnitude. This can be interpreted as increased light intensity filling the bands with a higher density of photo-generated charge carriers, leading to faster recombination. We do not yet have enough TRMC data to know whether the modest reduction in carrier lifetime is due to aging or inherent sample-to-sample variability, but nonetheless a carrier lifetime of ~30 ns is highly encouraging for an emerging absorber material that has undergone minimal optimization. To put this in context, carrier lifetimes reported for CZTS thin-film absorbers, which have been heavily researched for at least 2 decades, are only a few tens of ns.[71]

*First-principles computations*

To gain atomistic insights into the measured optoelectronic properties, we have performed first-principles calculations of the electronic structure and defect properties in $CaZn_2P_2$ using the screened hybrid functional of Heyd-Scuseria-Ernzerhof (HSE06).[72] The calculated electron band structure and density of states are shown in **Figure 5a**. The upper valence band is mainly of P character, while the lower conduction band is mainly of Ca and Zn characters. We see that the valence-band maximum (VBM) of $CaZn_2P_2$ is located at the Γ point, with the conduction-band minimum (CBM) located at M point. This results in an indirect band gap of 1.55 eV and a direct band gap of 1.89 eV at the Γ point. The calculated band gaps fall between the values reported in previous theoretical work on $CaZn_2P_2$ [42, 73]. Our experimental optical absorption and PL spectra agree well with the electronic structure reported here. We note that the lowest conduction band exhibits two additional extrema at A and L points which are quite close in energy to the extrema at Γ point. This increases the density of states at the direct band gap energy and will enhance the direct band gap optical transitions. This further confirm the observed room-temperature PL peak at 1.95 eV to be direct band gap emissions.

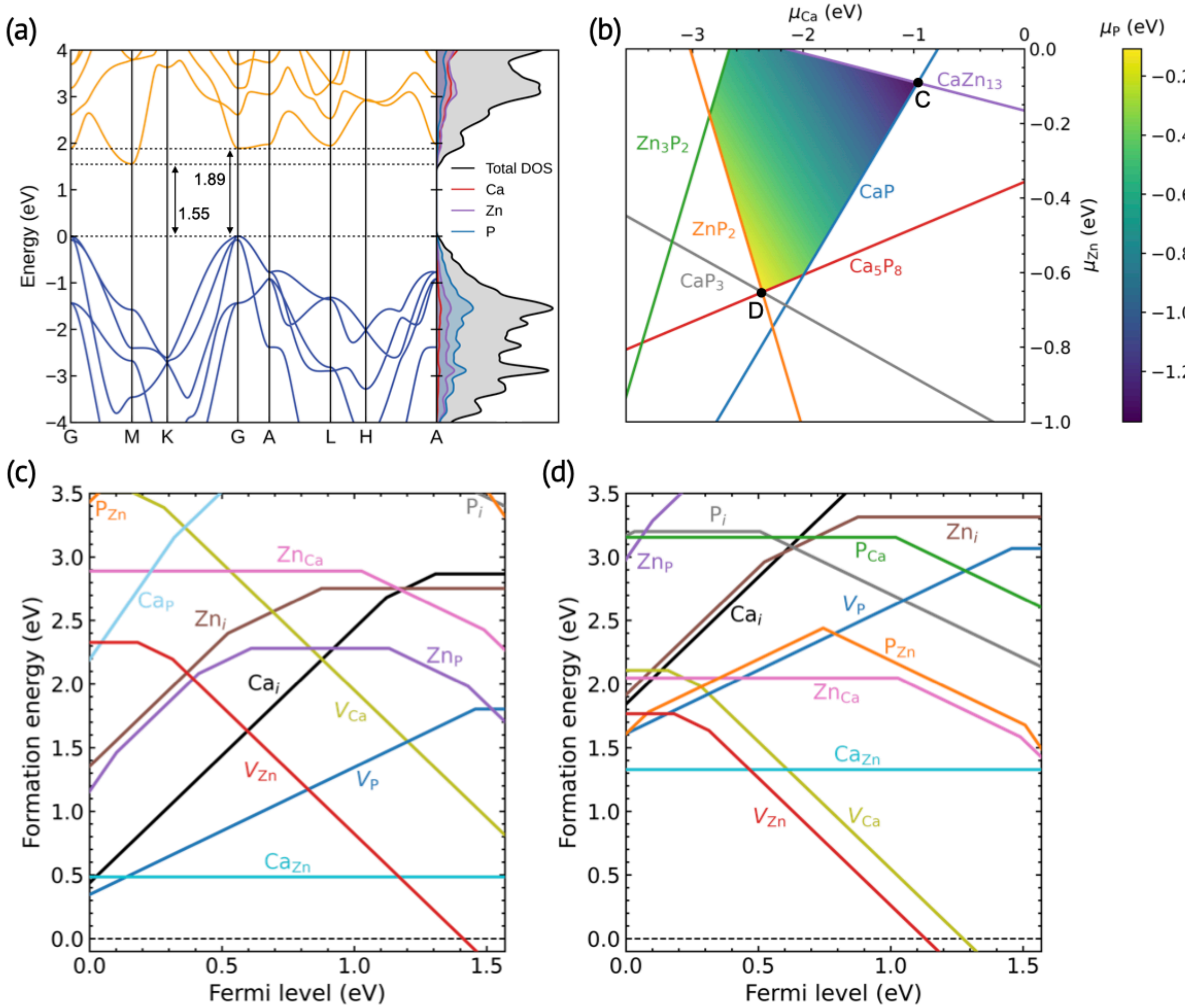

**Figure 5.** Electronic structure and defect properties in $CaZn_2P_2$ obtained from first-principles calculations based on the HSE06 functional. (a) Electron band structure with total and element-resolved density of states. (b) Chemical-potential stable region (colored area). (c)–(d) Defect formation energies as a function of Fermi level at the chemical-potential points C and D (which reflect P-poor and P-rich growth conditions, respectively).

The highest valence band and lowest conduction band of $CaZn_2P_2$ are overall dispersive, which would be beneficial to carrier transport. As documented in the Materials Project, the conductivity effective masses, deduced from the Boltzmann transport equation considering the upper valence bands and lower conduction bands, are 0.4–0.58 $m_0$ for holes and 0.25–1.29 $m_0$ for electrons.[74, 75] These values agree with a simple parabolic fit of the bands around the respective band extrema, resulting in 0.28–0.61 $m_0$ for holes and 0.14–1.04 $m_0$ for electrons; see also **Table S2**. The results suggest quite anisotropic electron transport in the quasi-layered $CaZn_2P_2$.

**Figures 5b** shows the computed stable region of $CaZn_2P_2$ in the elemental chemical potential space $(\mu_{\mathrm{Ca}}, \mu_{\mathrm{Zn}}, \mu_{\mathrm{P}})$. As can be seen, $CaZn_2P_2$ is thermodynamically stable against a series of binary compounds, including $CaZn_{13}$, CaP, $Ca_5P_8$, $CaP_3$, $ZnP_2$, and $Zn_3P_2$. Yet, $CaZn_2P_2$ has a wide stable region, indicating a large thermodynamic window for synthesizing phase-pure $CaZn_2P_2$. The allowed $\mu_{\mathrm{Ca}}$ and $\mu_{\mathrm{P}}$ ranges are large, while the $\mu_{\mathrm{Zn}}$ range is relatively narrow and restricted to between $-0.65$ and 0 eV, suggesting high Zn content is needed for $CaZn_2P_2$ to be stable under equilibrium growth conditions; here $\mu_{\mathrm{Zn}} = 0$ eV means that the Zn content is so high that pure Zn metal can form (see more details in Methods). Even though sputtering growth is a nonequilibrium process, the overall large region of thermodynamic stability helps to explain the wide compositional phase width where single-phase $CaZn_2P_2$ is observed in our combinatorial thin films. Furthermore, while some sub-stoichiometric Zn can be tolerated, we see amorphization of the films under Zn-poor conditions (*cf.* **Figure 3** and **Figure S1**). To reflect our experimental control of $PH_3$ partial pressure in the growth chamber, in **Figure 5b** we label two chemical-potential points C and D, which represent P-poor and P-rich growth conditions, respectively.

The calculated defect formation energies as a function of Fermi level under P-poor and P-rich conditions are shown in **Figures 5c** and **5d,** respectively. We find that $CaZn_2P_2$ shows favorable defect properties for a solar absorber. There are only a few intrinsic defects that can have low formation energy (say, less than 1 eV) and thus exist in significant concentrations. They are the vacancies ($V_{\mathrm{Ca}}$, $V_{\mathrm{Zn}}$, and $V_{\mathrm{P}}$), Ca-on-Zn antisite ($\mathrm{Ca_{Zn}}$), and Ca interstitial ($\mathrm{Ca}_i$). The $\mathrm{Ca_{Zn}}$ introduces no defect levels in the $CaZn_2P_2$ band gap, and all the three vacancies are shallow defects. The $\mathrm{Ca}_i$ has a relatively deep (+/2+) level, which could cause electron trapping, but its formation energy is only low enough under P-poor (i.e., Ca-rich) conditions and for Fermi levels close to the VBM (**Figures 5c**). On the other hand, all the other deep defects, such as $\mathrm{Zn_{Ca}}$, $\mathrm{P_{Zn}}$, and $\mathrm{Zn_P}$, have high formation energy. The absence of low-formation-energy, deep intrinsic defects, which could act as nonradiative carrier recombination centers limiting carrier lifetime,[76-79] suggests that $CaZn_2P_2$ has a high defect tolerance. This explains the long carrier lifetime measured in our $CaZn_2P_2$ thin films. Furthermore, the overall similar (and favorable) defect properties between $CaZn_2P_2$ and $BaCd_2P_2$,[46] and to a lesser extent even $SrZn_2N_2$,[80] suggests that a high defect tolerance could be a general feature across the entire $AM_2Pn_2$ ($Pn$ = pnictogens) family of materials.

Additionally, from **Figure 5c**, we find that under P-poor conditions, and in the absence of any impurities, $CaZn_2P_2$ would be an intrinsic material, in which the Fermi level would be pinned close to the intersection of the formation-energy lines of $V_{\mathrm{Zn}}$ and $V_{\mathrm{P}}$ and lies close to the mid gap. Under P-rich conditions (**Figure 5d**), $CaZn_2P_2$ is expected to be weakly p-type doped since the formation energy of the $V_{\mathrm{P}}$ donors is much higher than that of the $V_{\mathrm{Zn}}$ acceptors for most Fermi-level positions in the band gap; yet, the formation energy of $V_{\mathrm{Zn}}$ is quite high for Fermi level close to the VBM. This also suggests that under P-rich conditions, $CaZn_2P_2$ has a high potential to achieve p-type doping provided that a suitable acceptor dopant (possible candidates include Cu and K) is introduced. Compared to $BaCd_2P_2$, should be easier to achieve p-type doping in $CaZn_2P_2$, because the formation energy of compensating $V_P$ donors under P-rich conditions is higher in $CaZn_2P_2$ than in $BaCd_2P_2$. [46]

## Conclusion

$CaZn_2P_2$ is proposed as a high-performance solar absorber for tandem top cell applications. Crystalline thin films of $CaZn_2P_2$ were synthesized at low temperature using a reactive sputtering approach. We find that the films are semiconductors with a ~1.95 eV optical transition, as measured by UV-vis spectroscopy and photoluminescence, and confirmed by first-principles calculations using hybrid functionals. Using time-resolved microwave conductivity, we measure a carrier lifetime in $CaZn_2P_2$ thin films of up to 30 ns, confirming the promise of this new material as a candidate solar absorber. The long carrier lifetime is possible through favorable intrinsic defect properties of $CaZn_2P_2$, which shows absence of low-energy deep defects across the material's thermodynamic stability window. While XPS analysis showed that some surface degradation is observed in $CaZn_2P_2$, the bulk optoelectronic properties are stable for months in air and after exposure to $H_2O$. Between our experimental and theoretical data, we confirm that practically all the important properties for a tandem top cell absorber are favorable in $CaZn_2P_2$. Altogether, this work highlights the promise of $CaZn_2P_2$ thin films as solar absorbers and motivates future study on $CaZn_2P_2$ and related Zintl phase materials and devices.

## Acknowledgements

This work was authored in part by the National Renewable Energy Laboratory, operated by Alliance for Sustainable Energy, LLC, for the U.S. Department of Energy (DOE) under contract no. DE-AC36-08GO28308. Funding provided by the U.S. Department of Energy, Office of Science, Basic Energy Sciences, Division of Materials Science and Engineering, Physical Behavior of Materials program through the Office of Science Funding Opportunity Announcement (FOA) Number DE-FOA-0002676: Chemical and Materials Sciences to Advance Clean-Energy Technologies and Transform Manufacturing under award number DE-SC0023509. This research used resources of the National Energy Research Scientific Computing Center (NERSC), a DOE Office of Science User Facility supported by the Office of Science of the U.S. Department of Energy under contract no. DE-AC02-05CH11231 using NERSC award BES-ERCAP0023830. Thanks to Dr. Nicholas Strange for support with GIWAXS measurements. Use of the Stanford Synchrotron Radiation Lightsource, SLAC National Accelerator Laboratory, is supported by the U.S. Department of Energy, Office of Science, Office of Basic Energy Sciences under Contract No. DE-AC02-76SF00515. The authors acknowledge the use of facilities and instrumentation at the UC Irvine Materials Research Institute (IMRI), which is supported in part by the National Science Foundation through the UC Irvine Materials Research Science and Engineering Center (DMR-2011967). Thanks to Dr. Ich Tran for support with photoelectron spectroscopy data collection. This work was performed in part at the San Diego Nanotechnology Infrastructure (SDNI) of UC San Diego, a member of the National Nanotechnology Coordinated Infrastructure, which is supported by the National Science

Foundation (grant ECCS-2025752). The views expressed in the article do not necessarily represent the views of the DOE or the U.S. Government.

**Authorship Statement**

Conceptualization, S.Q. and S.R.B.; Methodology, S.Q., A.Z., and S.R.B.; Investigation, S.Q., Z.Y., O.G.R., J.M., G.E., S.D., A.P., M.R.H., G.K., and S.R.B.; Writing—original draft, S.Q. and S.R.B.; Writing—reviewing and editing, all coauthors; Funding Acquisition, J.L., K.K., D.P.F., A.Z., G.H., and S.R.B.; Supervision, J.L., K.K., D.P.F., G.H., and S.R.B.

**Conflicts of interest**

There are no conflicts to declare.

**Methods**

Experimental data management and visualization in this work were done using the COMBIgor package.[81] Crystal structures were generated using the VESTA software.[82] For some figures, the scientific color scale batlow was used to prevent visual distortion and ensure readability for people with color-vision deficiency.[83]

*Thin film synthesis*

*Warning: We strongly emphasize that $PH_3$ is toxic and pyrophoric and P deposits leftover in a growth chamber can spontaneously combust during venting and routine chamber service (part changes, cleaning, etc.). Thus, additional safety controls, including robust interlocking, hydride gas monitoring, pump/purge cycling, self-contained breathing apparatus, flame-retardant personal-protective equipment, exhaust abatement, and others, must be rigorously implemented from the onset for any growth chamber intended to utilize $PH_3$ or prepare phosphide samples. Details on the deposition system and safely handling $PH_3$ as a process gas is described in our previous work.*[57]

$CaZn_2P_2$ thin films were synthesized by radio frequency (RF) co-sputtering from metallic 50.8 mm diameter Ca and Zn targets in a mixed $PH_3$/Ar gas environment. The process gas was introduced into the chamber as a mixture of 2% $PH_3$ and 98% Ar flowing at 19.5 sccm. The chamber pressure was maintained at 5 mTorr by a throttled gate valve, corresponding to 0.1 mTorr of $PH_3$. While the growth chamber's base pressure was < $10^{-7}$ Torr, after every growth an $O_2$ purging step is carried out, prior to shuttling the sample to the load lock, to clear any unreacted P from the growth platen and $PH_3$ from the chamber, so some oxygen might be expected. Films were made with both combinatorial composition gradients, achieved by keeping the substrate stationary against the confocally oriented sputter cathodes, as well as with compositional uniformity, achieved by rotating the platen during growth. The substrate temperature ($T_{growth}$) was varied from 100°C – 400°C. Crystalline $CaZn_2P_2$ thin films were successfully grown on both conductive and non-conductive substrates, including aSiO2 (fused silica), borosilicate glass (Corning EXG), Si (100) orientation with native oxide), c-plane sapphire, and fluorinated tin oxide coated glass (FTO), with a deposition time of 2 hours yielding films approximately 500 nm thick.

*Bulk polycrystalline sample synthesis*

*Warning: The starting reagent, metallic Ca is air- and water-reactive and should be handled carefully in an inert atmosphere. At >400 °C inside the reaction ampoule, excessive vapor pressure of P or reaction of Ca with silica may compromise the silica ampoule resulting in shattering or explosion. The annealing steps must be conducted in a well-ventilated environment, such as in a fume hood. Placing ampoule into a preheated furnace is a hazardous procedure because sudden rise of temperature, and corresponding pressure of volatile P, may over pressurize ampoule leading to shattering. The amounts of sample in such experiments should be minimized and excessive protection measures are required – face-shield, thermal resistant gloves and lab coats, and long tongs at very minimum.*

$CaZn_2P_2$ bulk polycrystalline powders were synthesized at high temperatures from elements via solid state reaction. The elements, Ca (99.98%, Alfa Aesar), Zn (99.9%, Fisher Scientific), and red phosphorus (98.9%, Alfa Aesar) were weighed in stoichiometric 1:2:2 ratio inside Ar-filled glovebox and placed inside a carbonized silica ampoule with inner/outer diameters of 9/11 mm. The ampoule was then evacuated to approximately 35 µTorr pressure and sealed using a hydrogen-oxygen torch. The muffle furnace was preheated at 850°C, and sealed ampoule was placed inside a muffle furnace to minimize heating time. Ampoule was annealed at 850°C for 4 hours, after which it was allowed to cool naturally in turned off furnace. The ampoule was then opened inside glovebox under Ar atmosphere and the sample was grinded into fine powder, which was placed in another carbonized silica ampoule, evacuated and then sealed in the previously described manner. The ampoule was placed into a muffle furnace at room temperature, heated at 100°C/h rate to 1000°C, and annealed at that temperature for 72 hours after which the furnace was turned off and sample was allowed to cool naturally. After the second annealing the ampoule was open in an ambient atmosphere. Powder X-ray diffraction using with Rigaku Miniflex 600 diffractometer with a Cu-Kα radiation a Ni-$K_\beta$ filter confirms presence of the single-phase sample of $CaZn_2P_2$.

*Thin film characterization*

Combinatorial X-ray diffraction (XRD) measurements were conducted on a Bruker D8 diffractometer using Cu $K_\alpha$ radiation and a 2D detector. Patterns were integrated to generate an intensity vs. $2\theta$ pattern. Synchrotron grazing incidence wide angle X-ray scattering (GIWAXS) measurements were performed at beamline 11-3 at the Stanford Synchrotron Radiation Lightsource, SLAC National Accelerator Laboratory. The data were collected with a Rayonix 225 area detector using a wavelength of $\lambda$ = 0.97625 Å, a 3° incident angle, a 150 mm sample-to-detector distance, and a beam size of 50 µm vertical x 150 µm horizonal. The diffraction images were calibrated with a $LaB_6$ standard and integrated with the Nika SAS package. Integrated data were averaged from 5 frames of 15 seconds each. X-ray fluorescence (XRF) measurements were performed using a Rh anode at 50 keV and spectra were modelled as a $CaZn_2P_2$ layer with unknown composition and thickness on top of a Si or $aSiO_2$ substrate. Composition for films grown on borosilicate glass-based (i.e., Ca-containing) substrates is implied through calibrations.

Scanning electron micrographs (SEMs) of $CaZn_2P_2$ deposited atop fluorine-doped tin oxide (FTO) were collected on a Zeiss Sigma 500 microscope, operating at a 3.00 kV accelerating voltage, using an in-lens secondary electron detector. Micrographs were collected in coss-section and at a 45° tilt (isometric perspective). Cross section samples of the films were prepared by scoring the glass backside and cleaving using a straight edge. No conductive coating or additional modification was performed on the samples.

(Scanning) transmission electron microscopy ((S)TEM) high-angle annular dark-field (HAADF) and selected area electron diffraction (SAED) images were acquired with a Thermo Fisher Scientific Spectra 200 transmission electron microscope operating at an accelerating voltage of 200 keV. Specimens for TEM were prepared from deposited films via in situ focused ion beam lift-out methods [84] using an FEI Helios Nanolab 600i SEM/FIB DualBeam workstation. Chemical mapping was performed in the TEM using the Super-X energy-dispersive X-ray spectroscopy (EDS) system equipped with four windowless silicon drift detectors, allowing for high count rates and chemical sensitivity (down to 0.5–1 at%). The EDS data were quantified using a multi-polynomial parabolic background and absorption correction in Velox.

X-ray photoelectron spectroscopy (XPS) were performed using an AXIS-Supra by Kratos Analytical with Au and Cu calibration, using a Al K-alpha photon source. Etching was done using an Ar gas cluster ion source. The films measured were both aged in air for multiple weeks as well as stored in a nitrogen environment immediately after deposition, though these two conditions did not vary significantly. Survey and high-resolution scans of Ca 2p, Zn 2p, Zn 1s-P 2p, C 1s, and O 1s were performed on the films as-is (without cleaning) as well as after a sputtering. A 60 second etch time was deemed appropriate by using 10 second etch cycles and monitoring the strength of the C 1s signal. XPS peak fitting was done in CasaXPS using Tougaard and Shirley backgrounds. Lorenztian asymmetric lineshapes were used to fit the peaks, with scans to optimize the lineshape parameters. The identification of peaks for carbonates, phosphates, and oxides were based on existing literature that exclusively used Au and/or Cu instrument calibration. [85-90] Identification of the $CaZn_2P_2$ peaks was done based on the peaks which increased in intensity after sputtering (relative to other peaks within the same scan), based on the assumption that the total signal from the $CaZn_2P_2$ would increase relative to the other phases as the initial surface was removed.

Optical transmission ($T$) and reflection ($R$) spectra were collected in the ultraviolet and visible (UV-vis) spectral regions on a custom-built optical spectroscopy system. A blank substrate was measured as a perfect transmission standard immediately before the $CaZn_2P_2$ film. An Al mirror was similarly used as a perfect reflection standard. The optical absorption ($A$) was determined from the fact that $A+T+R = 100\%$. The absorption coefficient, $\alpha$, was determined by numerically fitting the R, T data using transfer matrix method, as discussed in the main text. Photoluminescence (PL) measurements were performed on a Renishaw inVia (Gloucestershire,

UK) PL/Raman microscope equipped with a 532 nm laser and 20× magnification objective lenses. 600 lines $mm^{-1}$ grating was used to direct scattering light from the sample to the CCD detector

Time-resolved microwave conductivity (TRMC) measurements were performed on uniform $CaZn_2P_2$ samples grown on $aSiO_2$ substrates sized to fit into a custom-built ~10 GHz microwave cavity. The change in reflected microwave power is monitored as a function of both time and fluence as the sample is pumped with a pulsed Nd:YAG laser. Extensive details regarding the approach and system can be found elsewhere.[91]

*Theory*

The first-principles calculations in this work were performed using the VASP code (v6.3.2) and the screened hybrid functional of Heyd–Scuseria–Ernzerhof (HSE06). [72, 92, 93] The standard VASP projector augmented wave (PAW) pseudopotentials (Ca_sv, Zn, and P; version of PBE5.4) were used, and the plane-wave energy cutoff for the electron wave functions was set to 400 eV. Using a $\Gamma$-centered 8×8×4 $\boldsymbol{k}$-point grid, the lattice constants of the $CaZn_2P_2$ $P\bar{3}m1$ unit cell are calculated to be: $a = b = 4.03$ Å and $c = 6.83$ Å, in good agreement with our and previous measurements.[47] Based on the HSE06-calculated lattice parameters, we created a 4×4×3 supercell (which contains 240 atoms) for simulating the intrinsic point defects in $CaZn_2P_2$. For the supercell containing a point defect, the atomic positions were fully relaxed using a $\Gamma$-only $\boldsymbol{k}$-point grid and a force convergence criterion of 0.01 eV/ Å. Spin polarization was properly included in all the defect calculations.

The formation energy of a defect (denoted as $D$ below) in the charge state $q$ was computed using the standard first-principles formalism [94, 95]

$$E_\mathrm{f}(D^q) = E_\mathrm{tot}(D^q) - E_\mathrm{tot}(\mathrm{bulk}) + \sum_i n_i(E_i + \mu_i) + qE_F + \Delta^q,$$

where $E_\mathrm{f}(D^q)$ and $E_\mathrm{tot}(\mathrm{bulk})$ are the total energies of the defect-containing and defect-free supercells, respectively. The term $\Delta^q$ is a finite-supercell-size correction to $E_\mathrm{tot}(V_\mathrm{P}^q)$, and was obtained using the extended Freysoldt-Neugebauer-Van de Walle (FNV) scheme [96, 97] and the calculated (electronic) dielectric constants ($\varepsilon_{xx} = \varepsilon_{yy} = 11.13$, $\varepsilon_{zz} = 10.68$). The defect formation energy depends on the Fermi-level position $E_F$ which is referenced to the valence-band maximum (VBM) and can vary from the VBM to the conduction-band minimum (CBM). The formation energy depends also on the chemical potentials of the elements ($\mu_i$) involved in forming the defect which is created by moving (taking) $n_i$ atoms to (from) the atomic reservoir. $\mu_i$ is referenced to $E_i$ which is the total energy per atom of the pure phase of the element, so $\mu_i = 0$ represents the limit in which the elemental phase starts to form. For $CaZn_2P_2$, the chemical potentials of Ca, Zn, and P, namely $\mu_\mathrm{Ca}$, $\mu_\mathrm{Zn}$, and $\mu_\mathrm{P}$ are limited to the stable region of $CaZn_2P_2$ in the elemental chemical potential space ($\mu_\mathrm{Ca}, \mu_\mathrm{Zn}, \mu_\mathrm{P}$). The chemical-potential stable region of $CaZn_2P_2$ was determined by

$$\mu_\mathrm{Ca} + 2\mu_\mathrm{Zn} + 2\mu_\mathrm{P} = \Delta H_f(\mathrm{CaZn_2P_2}) = -3.89 \text{ eV},$$

and the following relations:

$$\mu_\mathrm{Ca} < 0 \text{ eV},$$

$$\mu_{\mathrm{Zn}} < 0 \text{ eV},$$
$$\mu_{\mathrm{P}} < 0 \text{ eV},$$
$$\mu_{\mathrm{Ca}} + 11\mu_{\mathrm{Zn}} < \Delta H_{\mathrm{f}}(\mathrm{CaZn_{13}})\ ) = -2.15 \text{ eV},$$
$$\mu_{\mathrm{Ca}} + \mu_{\mathrm{P}} < \Delta H_{\mathrm{f}}(\mathrm{CaP}) = -2.34 \text{ eV},$$
$$5\mu_{\mathrm{Ca}} + 8\mu_{\mathrm{P}} < \Delta H_{\mathrm{f}}(\mathrm{Ca_5P_8}) = -12.71 \text{ eV},$$
$$\mu_{\mathrm{Ca}} + 3\mu_{\mathrm{P}} < \Delta H_{\mathrm{f}}(\mathrm{CaP_3}) = -2.69 \text{ eV},$$
$$\mu_{\mathrm{Zn}} + 2\mu_{\mathrm{P}} < \Delta H_{\mathrm{f}}(\mathrm{ZnP_2}) = -0.87 \text{ eV},$$
$$3\mu_{\mathrm{Zn}} + 2\mu_{\mathrm{P}} < \Delta H_{\mathrm{f}}(\mathrm{Zn_3P_2})\ ) = -1.22 \text{ eV},$$

where $\Delta H_{\mathrm{f}}$ denotes the formation enthalpy per formula. The $\Delta H_f$ values were obtained from HSE06 calculations (including structural relaxations). These relations ensure thermodynamic stability of $CaZn_2P_2$ by avoiding formation of the elemental and binary phases.

The theoretical optical absorption spectra was obtained by calculating the frequency dependent dielectric function in the independent-particle approximation.[98] The calculations were performed using the HSE06 functional and a $\Gamma$-centered 10×10×5 **k**-point grid. The (small) complex shift in the Kramers-Kronig transformation was set to $10^{-5}$. The calculated optical absorption spectra in Fig. 4(b) in the main text arises from direct transitions from valence to conduction bands and is averaged over the diagonal Cartesian components (i.e., $xx$, $yy$, and $zz$).

**Supplemental information for:**

# Low-Temperature Synthesis of Stable $CaZn_2P_2$ Zintl Phosphide Thin Films as Candidate Top Absorbers

Shaham Quadir[1*], Zhenkun Yuan[2], Guillermo Esparza[3], Sita Dugu[1], John Mangum[1], Andrew Pike[2], Muhammad Rubaiat Hasan[4], Gideon Kassa[2], Xiaoxin Wang[2], Yagmur Coban[2], Jifeng Liu[2], Kirill Kovnir[4,5], David P. Fenning[3], Obadiah G. Reid[1,6], Andriy Zakutayev[1], Geoffroy Hautier[2], Sage R. Bauers[1†]

[1]Materials Science Center, National Renewable Energy Laboratory, Golden, CO 80401, USA

[2]Thayer School of Engineering, Dartmouth College, Hanover, NH 03755, USA

[3] Department of Chemical and Nano Engineering, University of California, San Diego, La Jolla, CA 92093, USA

[4] Department of Chemistry, Iowa State University, Ames, IA 50011, USA

[5] Ames National Laboratory, U.S. Department of Energy, Ames, IA 50011, USA

[6] Renewable and Sustainable Energy Institute, University of Colorado, Boulder, Boulder, CO 80309, USA

[*]shaham.quadir@nrel.gov

[†]sage.bauers@nrel.gov

**Contents**

**Supplementary Figures**

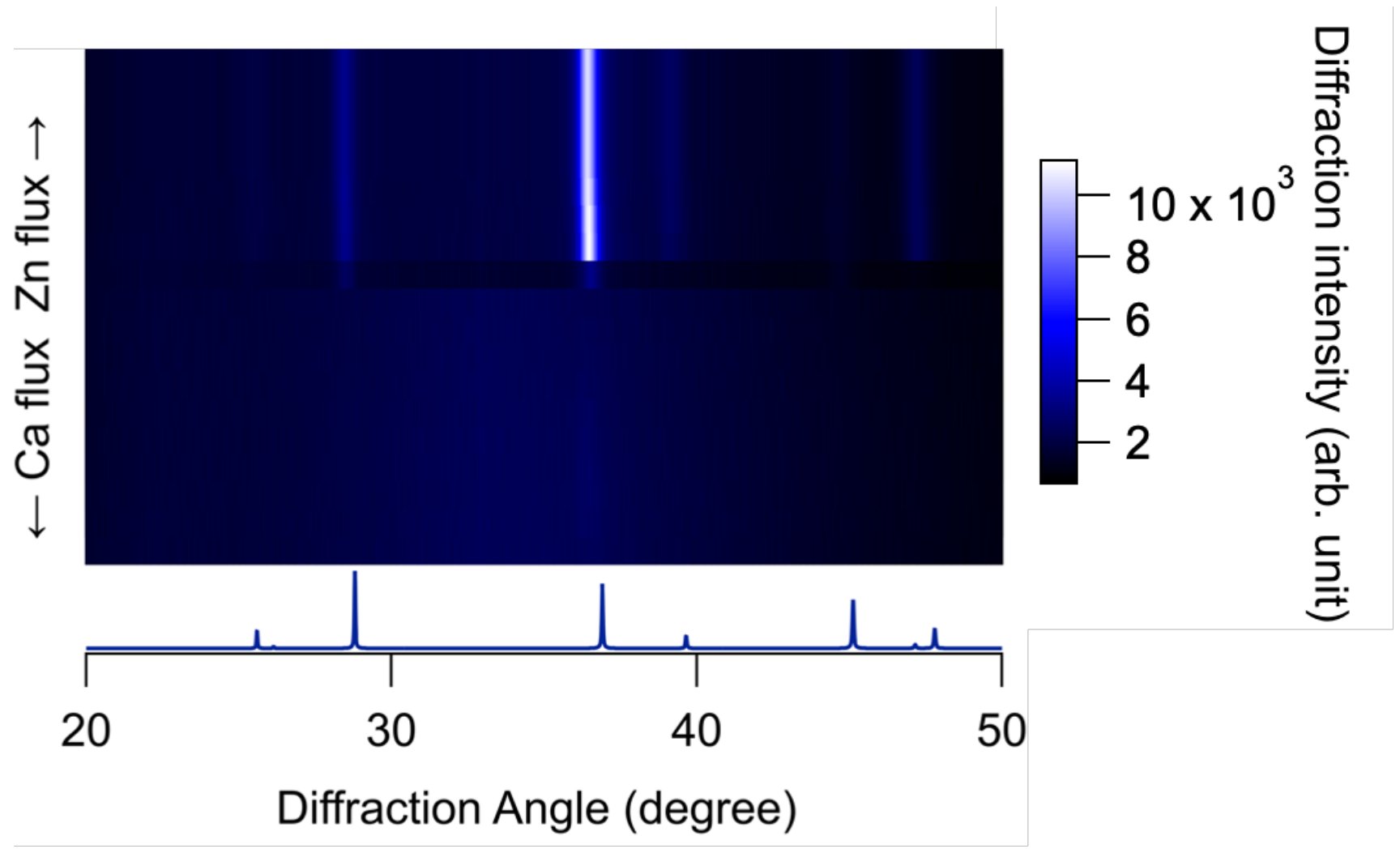

**Figure S1.** X-ray diffraction heatmap for $CaZn_2P_2$ films grown with high Ca flux ($T_{growth}$ = 200˚C)

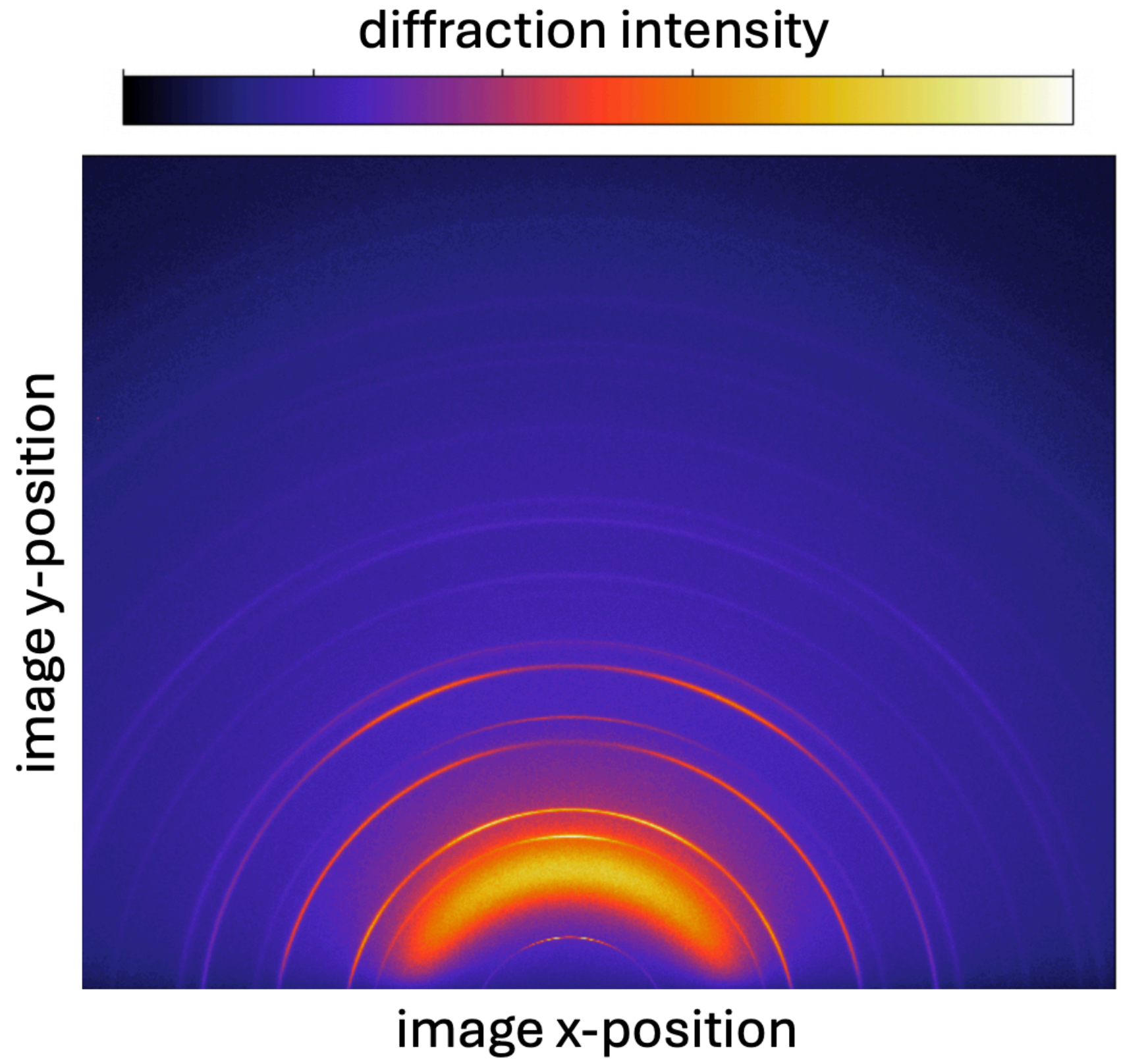

**Figure S2.** Synchrotron grazing incidence wide angle x-ray scattering pattern of a $CaZn_2P_2$ thin film. This is 1 of the 5 raw images that were summed, geometrically transformed, and integrated to create Figure 2d in the main text.

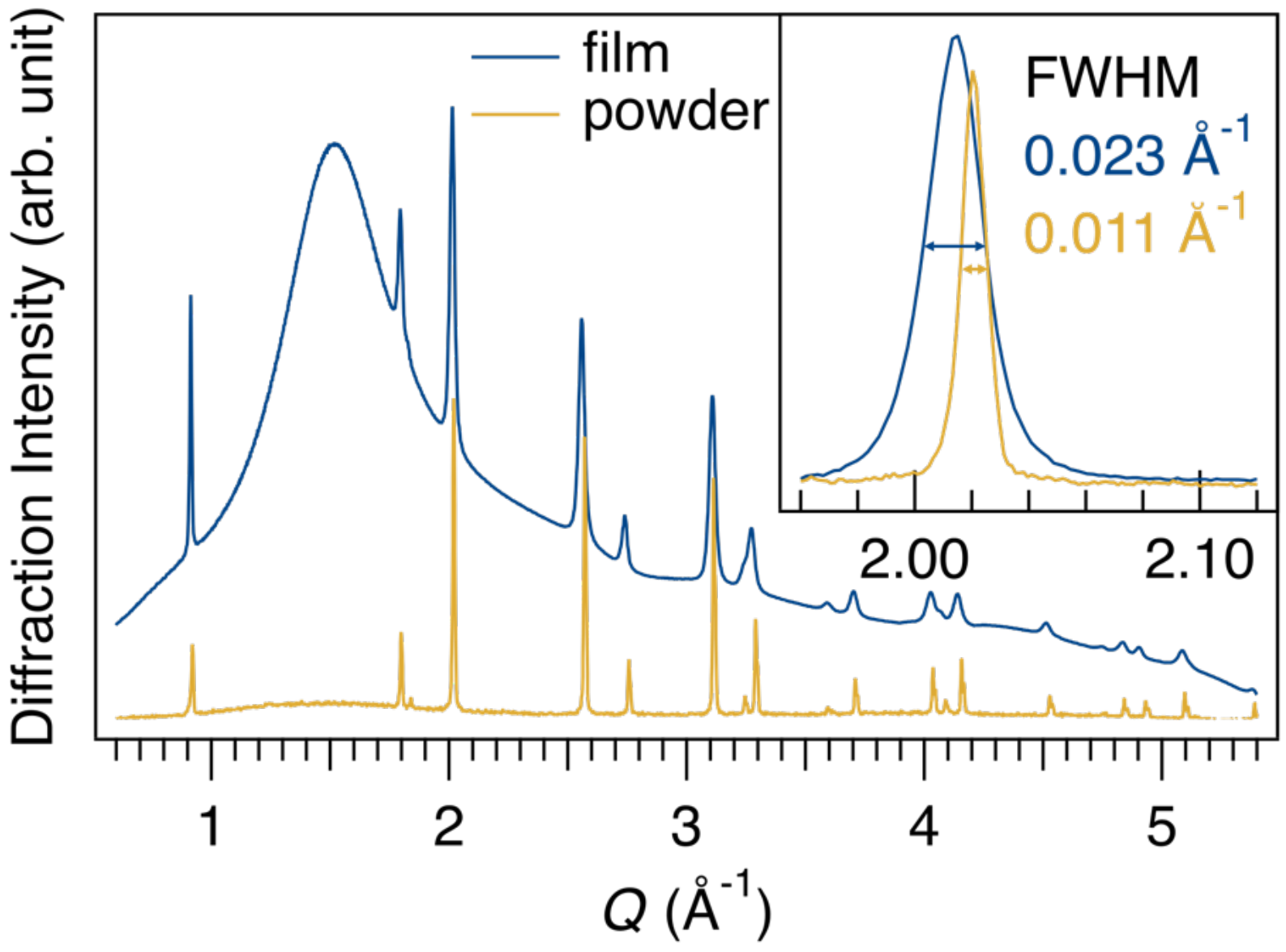

**Figure S3.** Comparison of X-ray diffraction patterns collected from a $CaZn_2P_2$ thin film and $CaZn_2P_2$ powder. Inset is a comparison of full width at half maximum (FWHM) for the (101) peak; note that the film data are background subtracted.

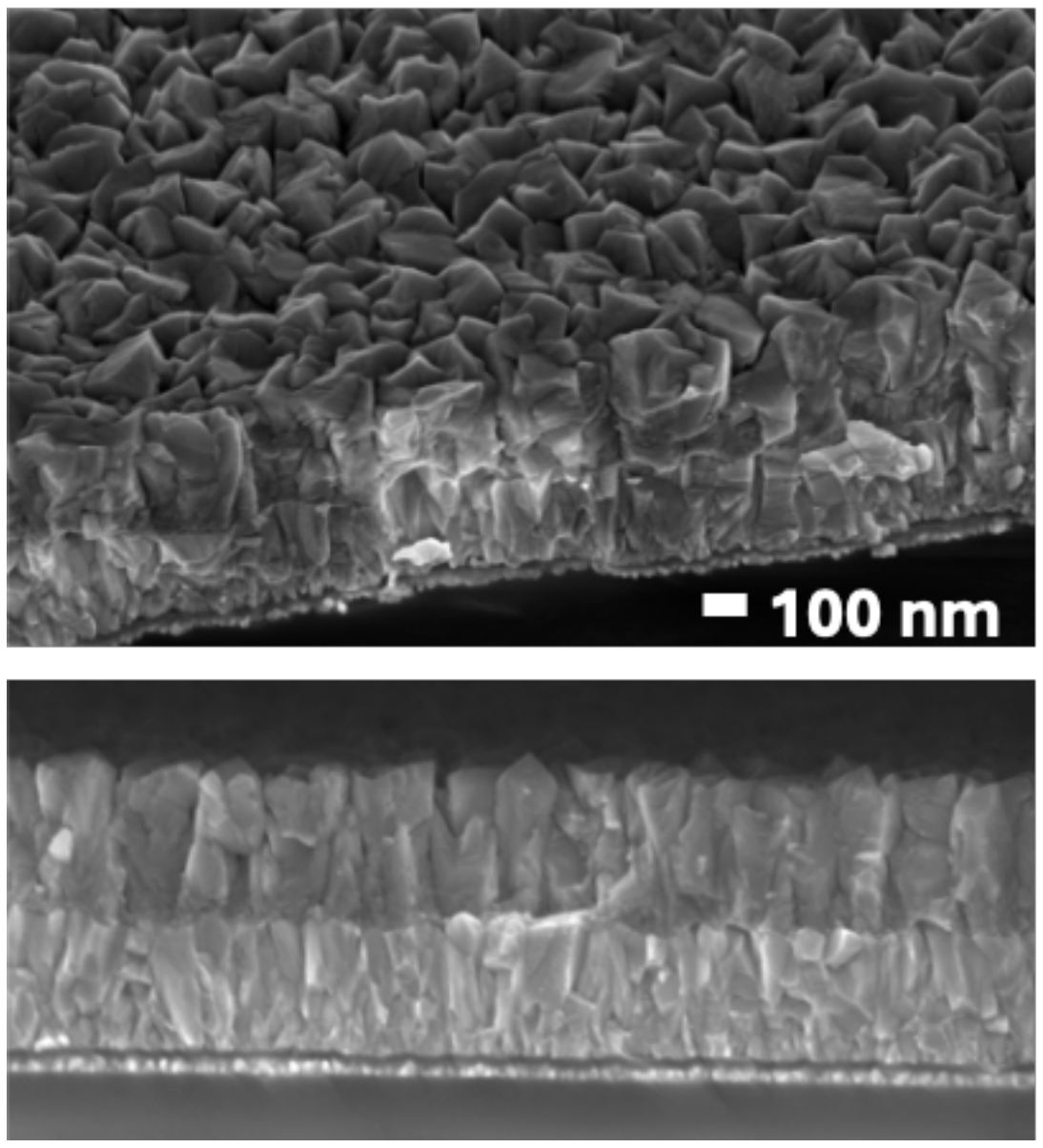

**Figure S4.** SEM isometric (top) and cross- -sectional view of a $CaZn_2P_2$ film grown on FTO. The lighter layer is the underlying FTO.

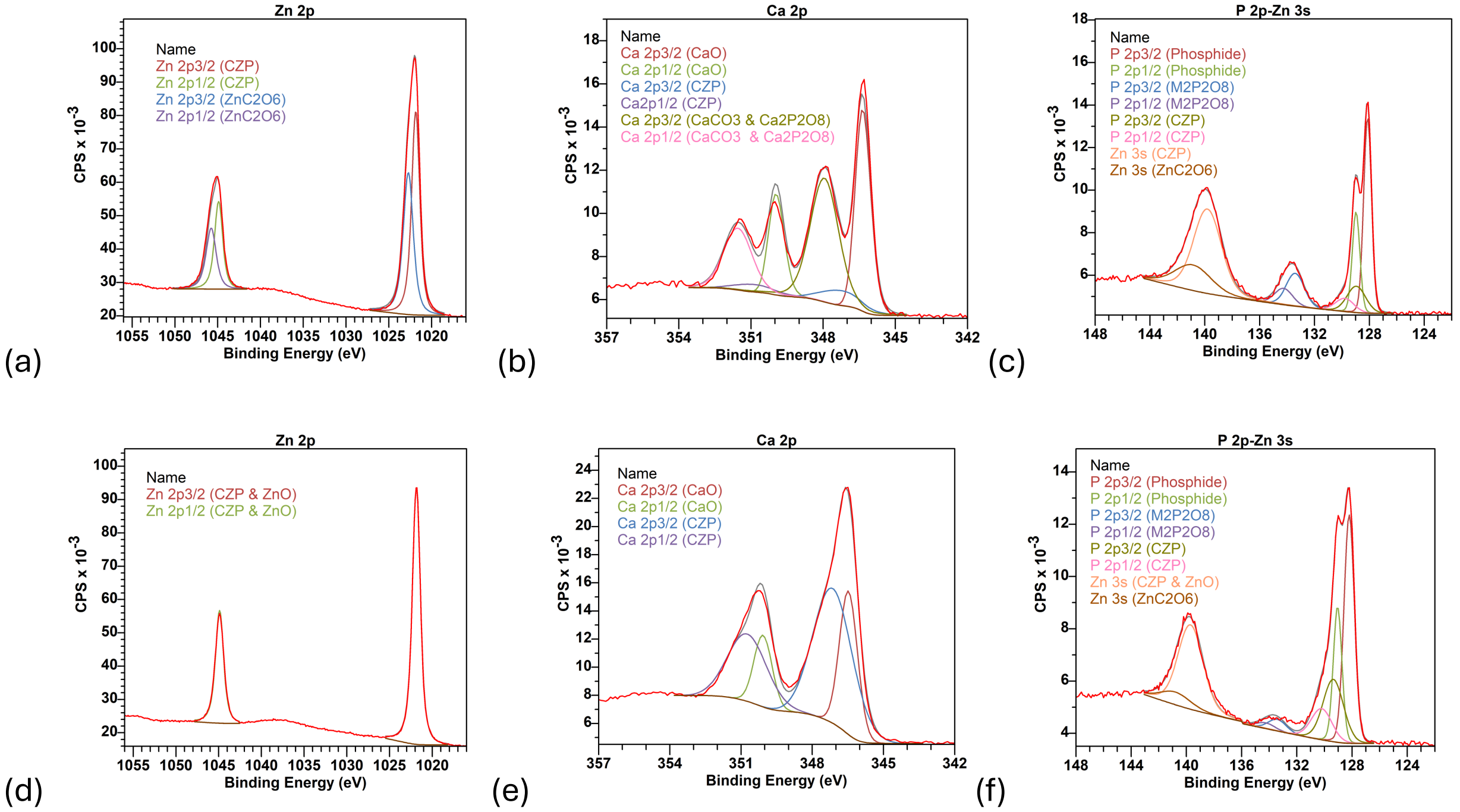

**Figure S5.** X-ray photoelectron (XPS) spectroscopy collected from a $CaZn_2P_2$ film grown on FTO. Panes (a), (b), (c) show spectra for Zn 2p, Ca 2p, and P 2p/Zn 3s electrons for films after exposure to air. Panes (d), (e), (f) show spectra for Zn 2p, Ca 2p, and P 2p/Zn 3s electrons for films after sputtering to remove adventitious C. Peak parameters are given in Tables S3 (neat) and S4 (etched).

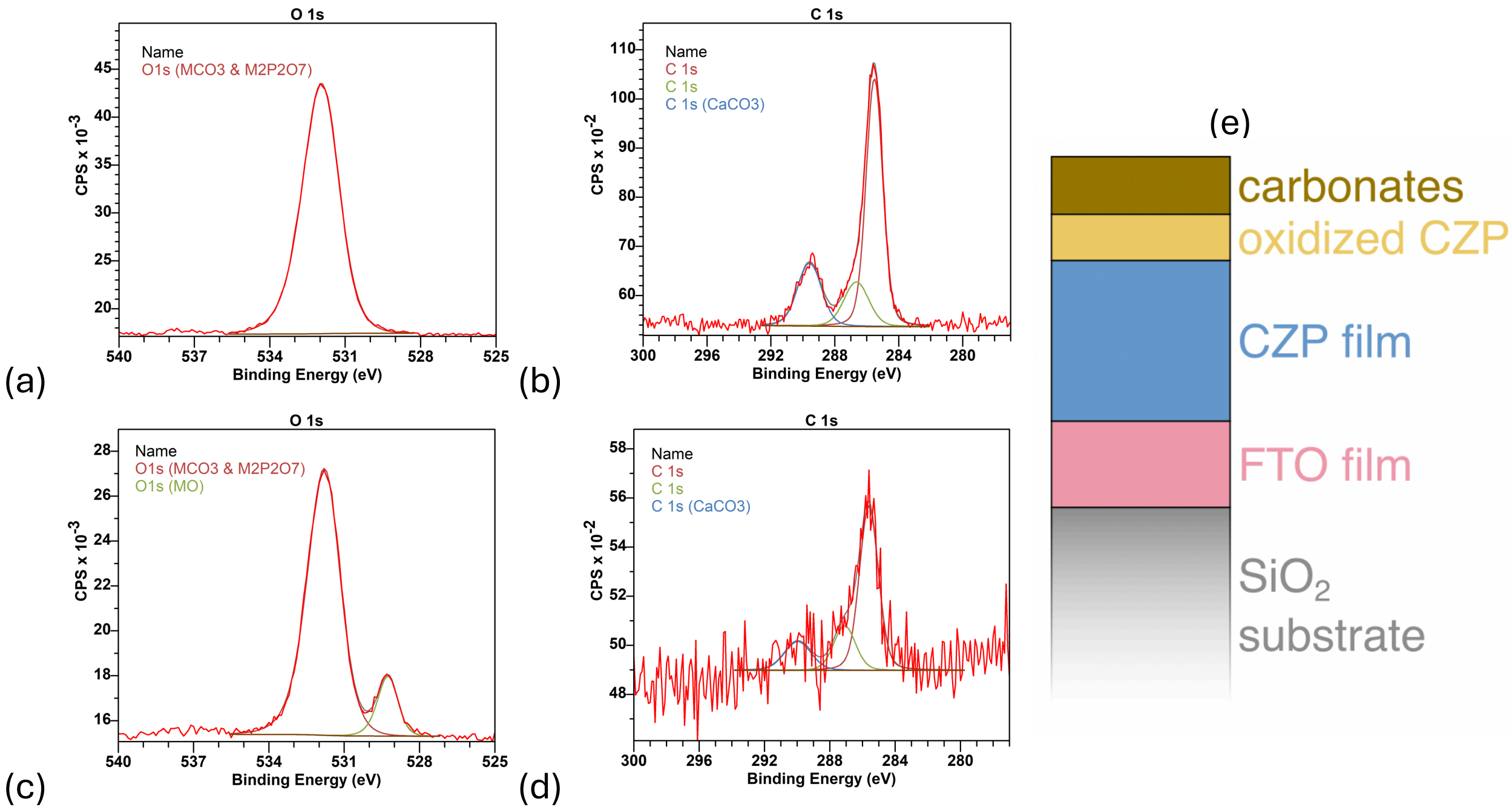

**Figure S6.** X-ray photoelectron (XPS) spectroscopy collected from a $CaZn_2P_2$ film grown on FTO. Panes (a) and (b) show XPS spectra for O 1s and C 1s electrons for neat films after exposure to air. Panes (c) and (d) show XPS spectra for O 1s and C 1s electrons for films after Ar-cluster sputtering to remove adventitious C. Pane (e) is the proposed layer stack for $CaZn_2P_2$ thin films grown on FTO substrates in a chamber purged with $O_2$ prior to exposing the films to atmosphere. Peak parameters are given in Tables S3 (neat) and S4 (etched).

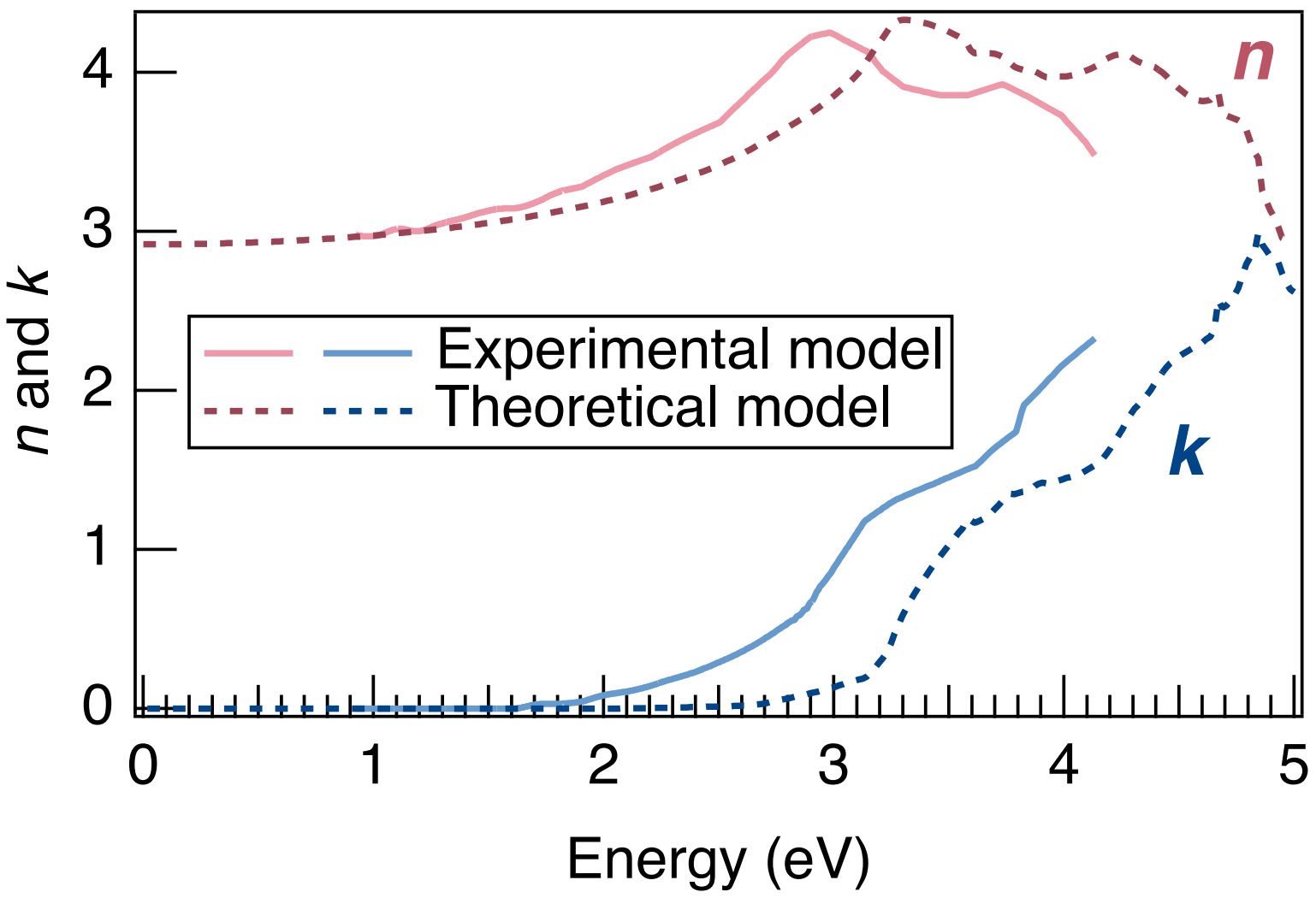

**Figure S7.** Experimentally modeled and theoretically calculated refractive index (n) and the extinction coefficient (k) of $CaZn_2P_2$ film.

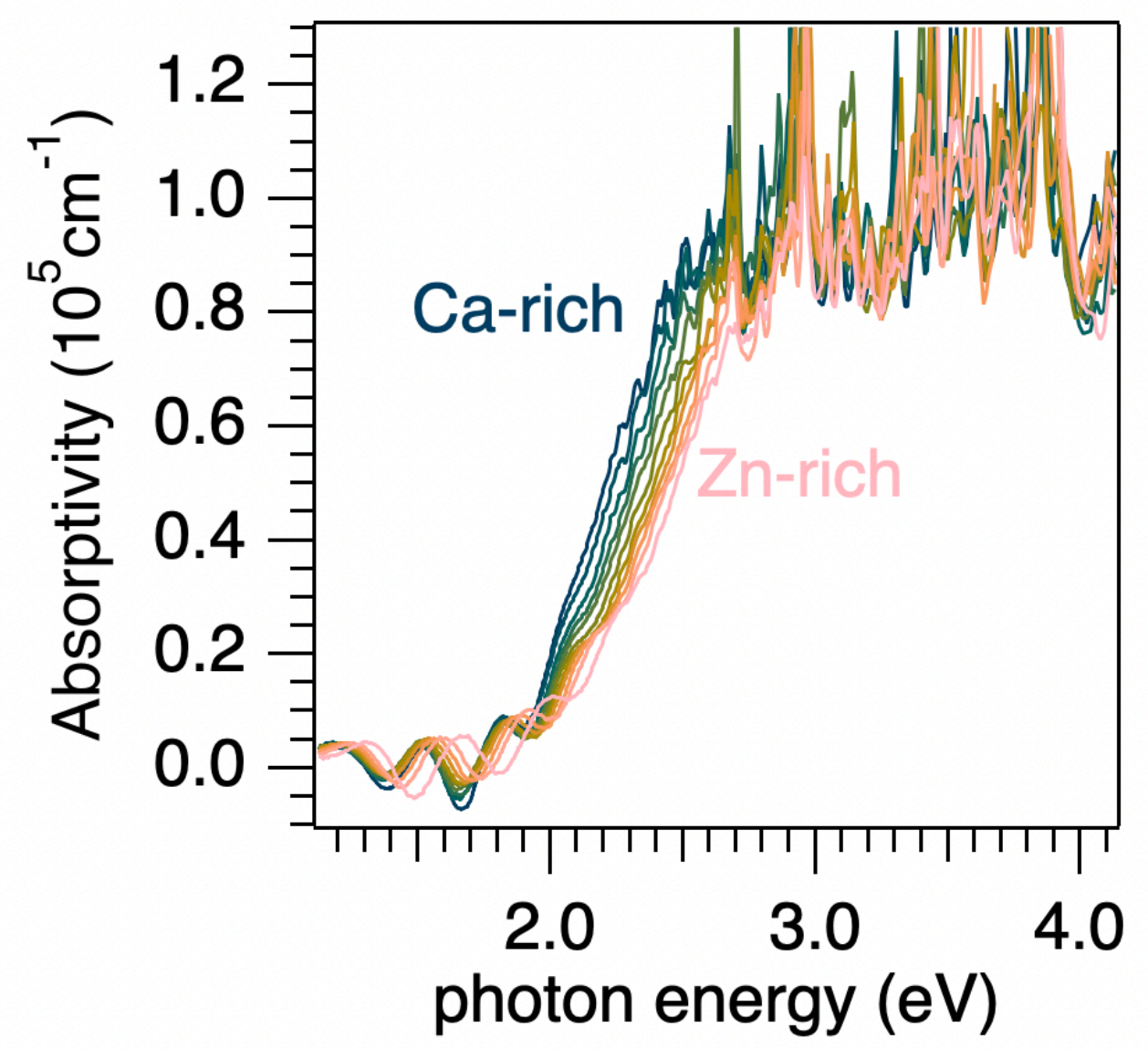

**Figure S8.** Absorption coefficient of a combinatorial $CaZn_2P_2$ film.

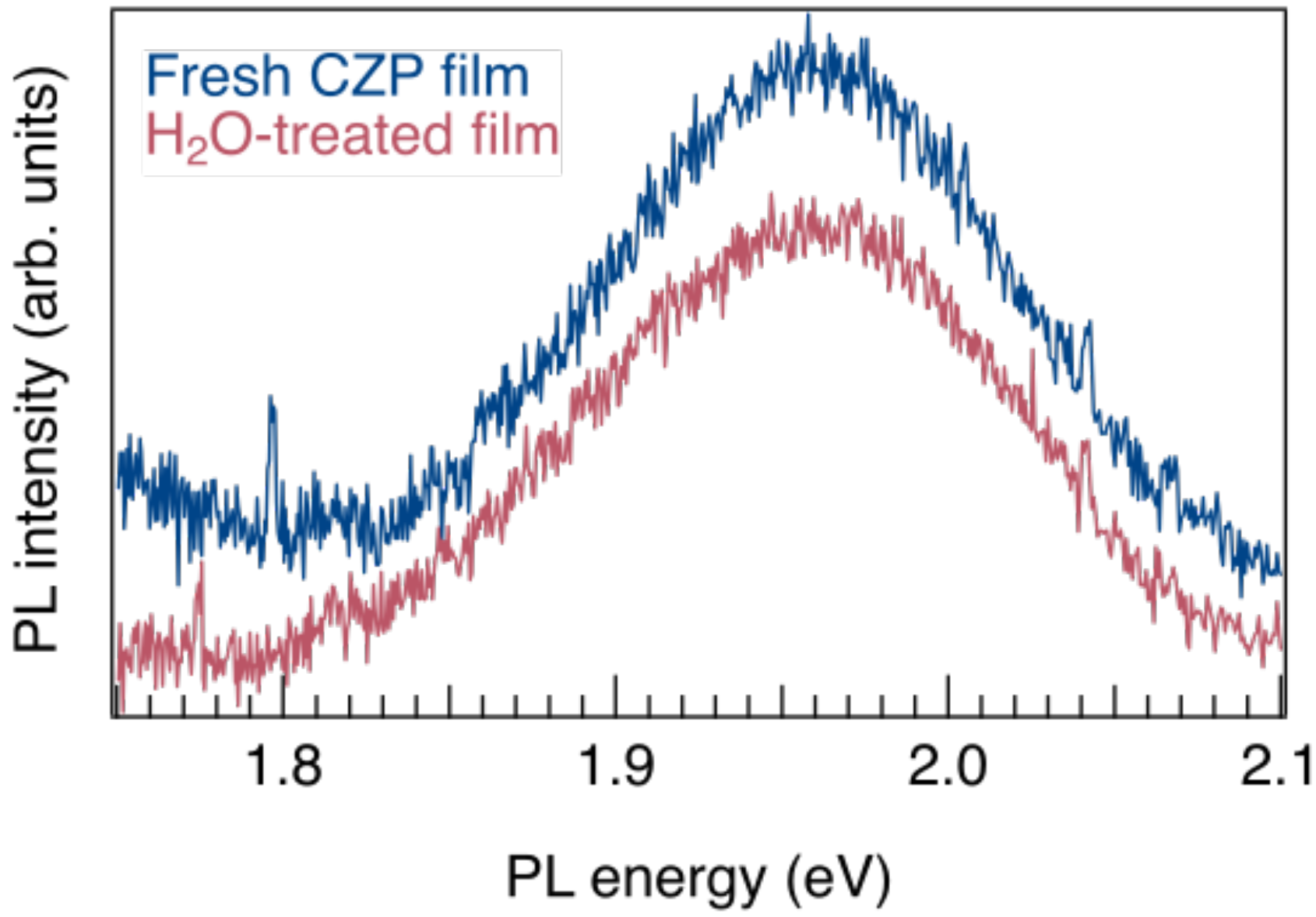

**Figure S9.** Photoluminescence of $CaZn_2P_2$, collected both before and after dipping in $H_2O$ for 90 seconds.

## Supplementary Tables

**Table S1.** Band gaps of $AM_2P_2$ compounds calculated using the HSE06 functional.

| Compound | $E_{g,\ direct}$ (eV) | $E_{g,\ fundamental}$ (eV) |
|---|---|---|
| $CaZn_2P_2$ | 1.89 | 1.55 |
| $CaCd_2P_2$ | 1.62 | 1.62 |
| $SrZn_2P_2$ | 1.74 | 1.50 |
| $SrCd_2P_2$ | 1.51 | 1.51 |
| $BaCd_2P_2$ | 1.45 | 1.45 |

**Table S2.** Effective masses of $CaZn_2P_2$ calculated using the HSE06 functional. The values are obtained from a parabolic fit the bands around the respective band extrema.

| System | Hole effective masses | Electron effective masses |
|---|---|---|
| $CaZn_2P_2$ | 0.35, 0.49 ($\Gamma - M$) | 1.04 (CBM$-\Gamma$) |
| | 0.28, 0.67 ($\Gamma - K$) | 0.14 (CBM$-K$) |
| | 0.61, 0.61 ($\Gamma - A$) | |

**Table S3.** XPS peak parameters for neat $CaZn_2P_2$ surface

| Name | Position | FWHM | Raw Area |
|---|---|---|---|
| Zn 2p3/2 ($CaZn_2P_2$) | 1021.85 | 1.11 | 87156.05 |
| Zn 2p1/2 ($CaZn_2P_2$) | 1044.89 | 1.11 | 37477.1 |
| Zn 2p3/2 ($ZnC_2O_6$) | 1022.71 | 1.36 | 74618.3 |
| Zn 2p1/2 ($ZnC_2O_6$) | 1045.75 | 1.36 | 32085.87 |
| O1s ($MCO_3$ & $M_2P_2O_7$) | 531.92 | 1.66 | 50594.1 |
| Ca 2p3/2 (CaO) | 346.44 | 0.74 | 7747.21 |
| Ca 2p1/2 (CaO) | 350.04 | 0.74 | 3873.61 |
| Ca 2p3/2 ($CaZn_2P_2$) | 347.49 | 1.94 | 1495.52 |
| Ca2p1/2 ($CaZn_2P_2$) | 351.09 | 1.94 | 747.76 |
| Ca 2p3/2 ($CaCO_3$ & $Ca_2P_2O_8$) | 347.95 | 1.38 | 9104.7 |
| Ca 2p1/2 ($CaCO_3$ & $Ca_2P_2O_8$) | 351.55 | 1.38 | 4552.35 |
| C 1s | 285.53 | 1.23 | 7095.14 |
| C 1s | 286.66 | 1.83 | 1909.16 |
| C 1s ($CaCO_3$) | 289.59 | 1.77 | 2596.28 |
| P 2p3/2 (Phosphide) | 128.14 | 0.67 | 6778.56 |
| P 2p1/2 (Phosphide) | 128.98 | 0.66 | 3389.28 |
| P 2p3/2 ($M_2P_2O_8$) | 133.41 | 1. 54 | 2604.43 |
| P 2p1/2 ($M_2P_2O_8$) | 134.25 | 1.54 | 1302.21 |
| P 2p3/2 ($CaZn_2P_2$) | 128.95 | 1.59 | 2162.21 |
| P 2p1/2 ($CaZn_2P_2$) | 129.79 | 1.59 | 1081.11 |
| Zn 3s ($CaZn_2P_2$) | 139.81 | 2.39 | 10937.43 |
| Zn 3s ($ZnC_2O_6$) | 140.86 | 3.13 | 4344.79 |

**Table S4.** XPS peak parameters for etched $CaZn_2P_2$ surface

| Name | Position | FWHM | Raw Area |
|---|---|---|---|
| Zn 2p3/2 ($CaZn_2P_2$ & ZnO) | 1021.85 | 1.1 | 109047.02 |
| Zn 2p1/2 ($CaZn_2P_2$ & ZnO) | 1044.89 | 1.1 | 47769.37 |
| O1s ($MCO_3$ & $M_2P_2O_7$) | 531.77 | 1.49 | 20733.07 |
| O1s (MO) | 529.24 | 0.91 | 2851.58 |
| Ca 2p3/2 (CaO) | 346.52 | 0.89 | 9770.57 |
| Ca 2p1/2 (CaO) | 350.12 | 0.89 | 4885.28 |
| Ca 2p3/2 ($CaZn_2P_2$) | 347.24 | 1.94 | 19950.81 |
| Ca 2p1/2 ($CaZn_2P_2$) | 350.84 | 1.94 | 9975.4 |
| C 1s | 285.61 | 1.36 | 1045.75 |
| C 1s | 287.09 | 1.58 | 328.13 |
| C 1s ($CaCO_3$) | 289.96 | 1.94 | 260.93 |
| P 2p3/2 (Phosphide) | 128.21 | 0.88 | 8443.89 |
| P 2p1/2 (Phosphide) | 129.05 | 0.74 | 4221.94 |
| P 2p3/2 ($M_2P_2O_8$) | 133.41 | 1.54 | 746.97 |
| P 2p1/2 ($M_2P_2O_8$) | 134.25 | 1.54 | 373.48 |
| P 2p3/2 ($CaZn_2P_2$) | 129.36 | 1.74 | 4540.73 |
| P 2p1/2 ($CaZn_2P_2$) | 130.2 | 1.74 | 2270.36 |
| Zn 3s ($CaZn_2P_2$ & ZnO) | 139.7 | 2.13 | 7940.25 |
| Zn 3s ($ZnC_2O_6$) | 140.79 | 2.58 | 1445.99 |